\documentclass[twoside,a4paper]{article}
\usepackage{amssymb}
\usepackage{xspace}
\usepackage{fullpage}
\usepackage{amsmath}
\usepackage{amsthm}
\usepackage{xcolor,svgcolor}
\usepackage{graphicx}
\usepackage{listings}
\usepackage{microtype}
\lstdefinestyle{slp}{
 basicstyle=\ttfamily\footnotesize,
 breaklines=true,
 breakatwhitespace=true,
 breakindent=0pt,
 string=[d]{;},
 keywordstyle=\color{blue},
 showstringspaces=false,
 xleftmargin=2pt, xrightmargin=2pt,
 numbers=none
}
\usepackage{ifpdf}
\ifpdf%
\usepackage{ae} %
\usepackage{aeguill}
\fi
\usepackage{embedfile}

\usepackage{hyperref}
\hypersetup{%
colorlinks,
citecolor=darkgreen,
filecolor=black,
linkcolor=darkred,
urlcolor=blue
}
\makeatletter
\def\mathcolor#1#{\@mathcolor{#1}}
\def\@mathcolor#1#2#3{%
 \protect\leavevmode
 \begingroup
 \color#1{#2}#3%
 \endgroup
}
\makeatother
\usepackage[nameinlink,capitalize,noabbrev]{cleveref}
\Crefname{proposition}{Proposition}{Propositions}
\crefname{proposition}{Proposition}{Propositions}
\crefname{equation}{Eq.}{Eqs.} %
\Crefname{equation}{Equation}{Equations}
\crefformat{footnote}{#2\footnotemark[#1]#3}
\newcommand{\SLP}{{\textsc{slp}}\xspace}
\newcommand{\SLPs}{{\textsc{slp}s}\xspace}
\newcommand{\lfrac}[2]{\ensuremath{{#1}/{#2}}}
\newcommand{\GO}[1]{\ensuremath{{O\!\left(\mathopen{}{#1}\right)\mathclose{}}}}
\newcommand{\firstdim}{\ensuremath{m}}
\newcommand{\seconddim}{\ensuremath{k}}
\newcommand{\thirddim}{\ensuremath{n}}
\newcommand{\Matrices}[2]{\ensuremath{{{#1}^{#2}}}}

\newcommand{\BDual}[1]{\ensuremath{{{\bigl({#1}\bigr)}^{\star}}}}
\newcommand{\matrixrank}[1]{\ensuremath{{{\textup{rank}\,{#1}}}}}
\newcommand{\LRPRepresentation}[3]{{\ensuremath{\left[{{#1};{#2};{#3}}\right]}}}

\newcommand{\row}[2]{#1_{#2,*}}
\newcommand{\col}[2]{#1_{*,#2}}
\newcommand{\vectorization}[1]{\ensuremath{\textup{Vect}{(#1)}}}%
\newcommand{\matrixsize}[2]{\ensuremath{{{#1}{\times}{#2}}}}
\newcommand{\mat}[1]{\ensuremath{\mathsf{#1}}}%
\newcommand{\MatrixProduct}[2]{\ensuremath{{{\mat{#1}}\cdot{\mat{#2}}}}}
\DeclareMathOperator{\Trace}{\ensuremath{\textup{Trace}}}
\newenvironment{smatrix}{\left[\begin{smallmatrix}}{\end{smallmatrix}\right]}
\newcommand{\Transpose}[1]{\ensuremath{{{#1}}^{\intercal}}}
\newcommand{\Inverse}[1]{\ensuremath{{#1}^{-1}}}
\newcommand{\InvTranspose}[1]{{{\mat{#1}}^{-\intercal}}}

\newcommand{\K}{\ensuremath{\mathrm{K}}}

\DeclareMathOperator{\tensorproduct}{\ensuremath{\otimes}}
\newcommand{\tensor}[1]{\ensuremath{\mathcal{#1}}}

\newcommand{\Contraction}[3]{\ensuremath{{{#1}\!\mid^{#2}_{#3}}}}
\newcommand{\LeftTensor}[1]{\ensuremath{{\mat{M}}_{#1}}}
\newcommand{\RightTensor}[1]{\ensuremath{{\mat{N}}_{#1}}}
\newcommand{\ProductTensor}[1]{\ensuremath{{\mat{O}}_{#1}}}
\newcommand{\FMMA}[4]{\ensuremath{{\langle{{#1}{\times}{#2}{\times}{#3}{:}{#4}}\rangle}}}
\newcommand{\DirectProduct}[2]{\ensuremath{{{#1}{\times}{#2}}}}
\newcommand{\SemiDirectProduct}[2]{\ensuremath{{{#1}{\rtimes}{#2}}}}
\newcommand{\SymmetricGroup}[1]{\ensuremath{{\mathfrak{S}_{#1}}}}
\newcommand{\CyclicGroup}[1]{\ensuremath{{\mathfrak{C}_{#1}}}}
\newcommand{\DihedralGroup}[1]{\ensuremath{{\mathfrak{D}_{#1}}}}
\DeclareMathOperator{\IsotropyComposition}{\circ}
\newcommand{\Isotropy}[1]{\ensuremath{\mathsf{#1}}}
\newcommand{\IsotropyAction}[2]{\ensuremath{{#1}\diamond{#2}}}
\def\triadone{brown}
\def\triadtwo{red}
\def\triadthree{blue}
\def\triadfour{green}
\def\triadfive{magenta}
\def\triadsix{gray}
\def\triadseven{violet}
\theoremstyle{plain}
\newtheorem{theorem}{Theorem}[section]

\newtheorem{lemma}{Lemma}[section]
\theoremstyle{definition}
\newtheorem{notation}{Notation}[section]

\newtheorem{definition}{Definition}[section]
\newtheorem{definitions}[definition]{Definitions}
\theoremstyle{remark}
\newtheorem{remark}{Remark}[section]

\newtheorem{example}{Example}[section]
\newtheorem{problem}{Question}[section]
\newcommand{\plinopt}{\href{https://github.com/jgdumas/plinopt}{\textsc{PLinOpt}}\xspace}
\newcommand{\plinoptdata}[1]{\href{https://github.com/jgdumas/plinopt/tree/main/data}{\url{#1}}}
\title{A non-commutative algorithm for multiplying~\(\matrixsize{4}{4}\) matrices using~\(48\) non-complex multiplications}
\author{%
 Jean-Guillaume Dumas, \\
 \href{mailto:Jean-Guillaume.Dumas@univ-grenoble-alpes.fr}{\texttt{Jean-Guillaume.Dumas@univ-grenoble-alpes.fr}}\\
 Univ.\ Grenoble Alpes, \textsc{cnrs}, Grenoble \textsc{inp} -- \textsc{uga}, \textsc{ljk},\\
 38000 Grenoble, France
 \and
 Cl\'ement Pernet, \\
 \href{mailto:clement.pernet@univ-grenoble-alpes.fr}{\texttt{Clement.Pernet@univ-grenoble-alpes.fr}} \\
 Grenoble \textsc{inp} -- \textsc{uga}, \textsc{cnrs}, Univ.\ Grenoble Alpes, \textsc{ljk},\\
 38000 Grenoble, France
 \and
 Alexandre Sedoglavic,\\
 \href{mailto:Alexandre.Sedoglavic@univ-lille.fr}{\texttt{Alexandre.Sedoglavic@univ-lille.fr}}\\
 Univ.\ Lille, \textsc{cnrs}, \textsc{umr} 9189 \textsc{cris}t\textsc{al},\\
 F-59000 Lille, France
}
\begin{document}
\maketitle
\begin{abstract}
The quest for non-commutative matrix multiplication algorithms over non-commutative rings in small dimensions has recently seen significant progress.
Specifically, the number of scalar multiplications required to multiply two~\(\matrixsize{4}{4}\) matrices was reduced in~\cite{Fawzi:2022aa} from~\(49\) (using two recursion levels of Strassen's algorithm) to~\(47\) in characteristic~\(2\), and more recently to~\(48\) in~\cite{alphaevolve} over the complex numbers.
\par
We propose an algorithm requiring~\(48\) multiplications that uses only rational coefficients,
thereby removing the requirement for complex-number arithmetic, and making this algorithm valid over
any ring except those of characteristic 2.
We also produce a straight-line program of this algorithm reducing
the number of additions and scalar multiplications, reaching a running time
of~\(\frac{347}{32}n^{2+\log_4{\!3}} +o\!\left({n^{2+\log_4{\!3}}}\right)\) operations, as well as
an alternative basis variant of 
it, leading to an algorithm running in~\({7n^{2+\log_4{\!3}} +o\!\left({n^{2+\log_4{\!3}}}\right)}\)
operations over any ring containing an inverse of~\(2\).
\par
Similarly, the number of scalar multiplications required to multiply a~\(\matrixsize{3}{4}\) matrix by a~\(\matrixsize{4}{7}\) matrix was reduced from~\(66\) in~\cite{Smirnov:2021aa} to~\(63\) in~\cite{alphaevolve} by an algorithm over complex numbers.
Using the same techniques, we propose an equivalent algorithm in~\(63\) multiplications using only rational coefficients.
In both cases the rational algorithm is obtained by identifying an isotropy that projects the
previously known complex-valued decomposition onto the field of rational numbers.
\end{abstract}
\section{Introduction}
Volker Strassen introduced in~\cite{strassen:1969} the first sub-cubic time algorithm for matrix multiplication, achieving a complexity bound of~\(\GO{n^{2.81}}\) using~\(7\) recursive multiplications.
\par
Recently, an algorithm requiring only~\(48\) scalar multiplications to multiply two~\(\matrixsize{4}{4}\) matrices over the field of complex numbers was discovered by means of a pipeline of large language models orchestrated by an evolutionary coding agent~\cite{alphaevolve}\footnote{A previous similar result was also announced in~\cite{Kaporin:2024ab}.}.
\par
We present here an \emph{equivalent variant} of this algorithm that relies solely on rational coefficients, completely avoiding complex values.
To achieve this, we consider the \emph{tensor decomposition representation} encoding the original algorithm and apply an \emph{isotropy} that maps it to another decomposition of necessarily the same \emph{tensor rank}, but relying exclusively on rational numbers.
\par
The next section outlines the mathematical framework that enables this transformation.
\Cref{sec:TheAlgorithm} presents the new rational tensor decomposition in a \emph{trilinear representation} alongside the specific isotropy used to obtain it and the stabilizer of the considered orbits.
\par
To convert this new tensor decomposition into a practical algorithm, we review the \textsc{lrp} (\emph{left-right-product}) representation of the tensor in~\Cref{sec:LRPrepresentation}.
We then demonstrate in~\Cref{sec:NumericalSchemeAndComplexity} how to transform this representation into a straight-line program, yielding a refined complexity bound:~\({\frac{347}{32}n^{2+\log_4{\!3}}-\frac{315}{32}n^{2}}\) that is approximately~\({10.84375n^{2.7925}}\).
In~\Cref{sec:AlternativeBasis}, we apply alternative basis decomposition~\cite{Karstadt:2017aa,Beniamini:2019aa} to achieve theoretically faster variants, reducing the complexity bound to~\(7n^{2+\log_4{\!3}}+o\!\left(n^{2+\log_4{\!3}}\right)\approx{7n^{2.7925}}\).
Finally,~\Cref{sec:347} applies the same techniques to present a~\(63\)-multiplication algorithm with rational coefficients for multiplying a~\(\matrixsize{3}{4}\) matrix by a~\(\matrixsize{4}{7}\) matrix.
\section{Tensor decomposition encoding matrix product}\label{sec:GeneralFramework}
We represent matrix multiplication algorithms through the tensor decomposition of their corresponding bilinmaps.
While a more comprehensive discussion of this framework is available in~\cite{Landsberg:2016ab}, we provide an overview here, using Strassen’s algorithm as a primary illustrative example.
\par
As a running example, consider three matrices~\({\mat{A},\mat{B},\mat{C}}\) defined as follows:
\begin{equation}
 \label{eq:InputMatricesABC}
 \mat{A}=\begin{bmatrix}
 {a_{11}}&{a_{12}}\\
 {a_{21}}&{a_{22}}
 \end{bmatrix},\quad
 \mat{B}= \begin{bmatrix}
 {b_{11}}&{b_{12}}\\
 {b_{21}}&{b_{22}}
 \end{bmatrix},\quad
 \mat{C}=\begin{bmatrix}
 {c_{11}}&{c_{12}}\\
 {c_{21}}&{c_{22}}
 \end{bmatrix}.
\end{equation}
The~\({\matrixsize{2}{2}}\) matrix product~\({\mat{C}=\MatrixProduct{\mat{A}}{\mat{B}}}\) can be computed using Strassen's algorithm that employs the following sequence of seven scalar multiplications:
\begin{equation}\label{eq:StrassenMultiplicationAlgorithm}
\begin{array}{ll}
\mathcolor{\triadone}{\rho_{1}}\leftarrow{\mathcolor{\triadone}{a_{11}}(\mathcolor{\triadone}{b_{12}-b_{22}})},
&
\mathcolor{\triadfour}{\rho_{4}}\leftarrow{(\mathcolor{\triadfour}{a_{12}-a_{22}})(\mathcolor{\triadfour}{b_{21}+b_{22}})},
\\
\mathcolor{\triadtwo}{\rho_{2}}\leftarrow{(\mathcolor{\triadtwo}{a_{11}+a_{12}})\mathcolor{\triadtwo}{b_{22}}},
&
\mathcolor{\triadfive}{\rho_{5}}\leftarrow{(\mathcolor{\triadfive}{a_{11}+a_{22}})(\mathcolor{\triadfive}{b_{11}+b_{22}})},
\\
\mathcolor{\triadthree}{\rho_{3}}\leftarrow{(\mathcolor{\triadthree}{a_{21}+a_{22}}) \mathcolor{\triadthree}{b_{11}}},
&
\mathcolor{\triadseven}{\rho_{7}}\leftarrow{(\mathcolor{\triadseven}{a_{21}-a_{11}})(\mathcolor{\triadseven}{b_{11}+b_{12}})},
\\
\mathcolor{\triadsix}{\rho_{6}}\leftarrow{\mathcolor{\triadsix}{a_{22}}(\mathcolor{\triadsix}{b_{21}-b_{11}})},
\\[\medskipamount]
\multicolumn{2}{c}{
 \begin{bmatrix} c_{11} &c_{12} \\ c_{21} &c_{22} \end{bmatrix}
 =
 \begin{bmatrix}
 \mathcolor{\triadfive}{\rho_{5}} + \mathcolor{\triadfour}{\rho_{4}} - \mathcolor{\triadtwo}{\rho_{2}} + \mathcolor{\triadsix}{\rho_{6}} &
 \mathcolor{\triadtwo}{\rho_{2}} + \mathcolor{\triadone}{\rho_{1}} \\
 \mathcolor{\triadsix}{\rho_{6}} + \mathcolor{\triadthree}{\rho_{3}}&
 \mathcolor{\triadfive}{\rho_{5}} + \mathcolor{\triadseven}{\rho_{7}} + \mathcolor{\triadone}{\rho_{1}}- \mathcolor{\triadthree}{\rho_{3}}
 \end{bmatrix}.}
\end{array}
\end{equation}
This straight-line program (a.k.a.~\SLP) encodes the following bilinear map over a field~\(\K\):
\begin{equation}\label{eq:mxnTimesnxp}
\beta_{\textsc{mm}}:
\begin{array}[t]{ccl}
\Matrices{\K}{\matrixsize{\firstdim}{\seconddim}}\times\Matrices{\K}{\matrixsize{\seconddim}{\thirddim}}&\rightarrow&\Matrices{\K}{\matrixsize{\firstdim}{\thirddim}},\\
(\mat{A},\mat{B})&\mapsto&\MatrixProduct{A}{B}.
\end{array}
\end{equation}
In the context of Strassen's algorithm, the dimensions~\({\firstdim,\seconddim,\thirddim}\) are all equal to~\(2\).
These indices are maintained throughout this section to clearly distinguish the different spaces (e.g.~\(\K^{\matrixsize{m}{k}}\) the space of~\(\matrixsize{m}{k}\) matrices with coefficients in~\(\K\)) involved in the following sections.
\subsection{Matrix multiplication encoded by tensor decomposition}\label{sec:tensordecomp}
Following the formalism in~\cite{Landsberg:2016ab}, we express the bilinear map of the matrix product as a tensor decomposition.
This offers a structured framework for representing associated algorithms and analyzing their inherent symmetries.
\par
Strassen's algorithm~(\Cref{eq:StrassenMultiplicationAlgorithm}) is encoded as the following matrix multiplication tensor decomposition~\FMMA{2}{2}{2}{7}, which is expressed as a sum of seven rank-one tensors~\({\LeftTensor{i}}\tensorproduct{\RightTensor{i}}\tensorproduct{\ProductTensor{i}}\):
\begin{equation}\label{eq:StrassenTensor}
\begin{array}{r}
\FMMA{2}{2}{2}{7}=\sum_{i=1}^{7}{\LeftTensor{i}}\tensorproduct{\RightTensor{i}}\tensorproduct{\ProductTensor{i}}=
\mathcolor{\triadfive}{{\begin{bmatrix}1&0\\0&1\end{bmatrix}}\tensorproduct{\begin{bmatrix}1&0\\0&1\end{bmatrix}}\tensorproduct\begin{bmatrix}1&0\\0&1\\\end{bmatrix}}
\\[\bigskipamount]
+
\mathcolor{\triadfour}{\begin{bmatrix}0&1\\0&-1\\\end{bmatrix}\tensorproduct\begin{bmatrix}0&0\\1&1\\\end{bmatrix}\tensorproduct\begin{bmatrix}1&0\\0&0\\\end{bmatrix}}
+
\mathcolor{\triadseven}{\begin{bmatrix}-1&0\\1&0\\\end{bmatrix}\tensorproduct\begin{bmatrix}1&1\\0&0\\\end{bmatrix}\tensorproduct\begin{bmatrix}0&0\\0&1\\\end{bmatrix}}
\\[\bigskipamount]
+
\mathcolor{\triadtwo}{\begin{bmatrix}1&1\\0&0\\\end{bmatrix}\tensorproduct\begin{bmatrix}0&0\\0&1\\\end{bmatrix}\tensorproduct\begin{bmatrix}-1&0\\1&0\\\end{bmatrix}}
+
\mathcolor{\triadone}{\begin{bmatrix}1&0\\0&0\\\end{bmatrix}\tensorproduct\begin{bmatrix}0&1\\0&-1\\\end{bmatrix}\tensorproduct\begin{bmatrix}0&0\\1&1\\\end{bmatrix}}
\\[\bigskipamount]
+
\mathcolor{\triadsix}{\begin{bmatrix}0&0\\0&1\\\end{bmatrix}\tensorproduct\begin{bmatrix}-1&0\\1&0\\\end{bmatrix}\tensorproduct\begin{bmatrix}1&1\\0&0\\\end{bmatrix}}
+
\mathcolor{\triadthree}{\begin{bmatrix}0&0\\1&1\\\end{bmatrix}\tensorproduct\begin{bmatrix}1&0\\0&0\\\end{bmatrix}\tensorproduct\begin{bmatrix}0&1\\0&-1\end{bmatrix}}
\end{array}
\end{equation}
in~\({\BDual{{\K}^{\matrixsize{\firstdim}{\seconddim}}}\tensorproduct{\BDual{{\K}^{\matrixsize{\seconddim}{\thirddim}}}}\tensorproduct{{\K}^{\matrixsize{\firstdim}{\thirddim}}}}\) (with~\({{\firstdim}={\seconddim}={\thirddim}={2}}\) in~\Cref{eq:StrassenTensor}).
\begin{definitions}\label{def:RankAndType}
Let~\(\tensor{T}\) be a tensor decomposition~\({\sum_{i=0}^{r}{{\mat{T}_{i1}}\tensorproduct{\mat{T}_{i2}}\tensorproduct{\mat{T}_{i3}}}}\) where each summand is a \emph{rank-one tensor}:
\begin{itemize}
\item the integer~\(r\) denotes the \emph{tensor rank} of the tensor decomposition~\(\tensor{T}\);
\item the ordered list~\({{[{(\matrixrank{\mat{T}_{ij}})}_{j={1}\ldots{3}}]}_{i={1}\ldots{r}}}\) defines the \emph{type} of tensor decomposition where~(\(\matrixrank{\mat{M}}\) denotes the classical rank of the matrix~\(\mat{M}\)).
The type can be compactly summarized as the multivariate polynomial~\({\sum_{i=0}^{r}{X^\matrixrank{\mat{T}_{i1}}}{Y^\matrixrank{\mat{T}_{i2}}}{Z^\matrixrank{\mat{T}_{i3}}}}\).
\end{itemize}
\end{definitions}
For the Strassen tensor decomposition, the \emph{tensor rank} is~\(7\) and its type is~\({X^{2}Y^{2}Z^{2}+6XYZ}\).
\par
The following relationship between the classical trace of the linear operators and the tensor representations we introduced above is central to the developments in~\Cref{sec:Isotropies} and~\Cref{sec:TheAlgorithm}.
The Frobenius inner product~\({\langle\ \rangle}\) of a matrix~\(\mat{O}\) with a matrix product~\(\MatrixProduct{M}{N}\) defines a trilinear form~\({\Trace(\MatrixProduct{\Transpose{O}}{\MatrixProduct{M}{N}})}\) as follows:
\begin{equation}\label{eq:TrilinearForm}
\Contraction{\tensor{S}}{}{3}:
\begin{array}[t]{ccc}
{\K}^{\matrixsize{\firstdim}{\seconddim}}\times{{\K}^{\matrixsize{\seconddim}{\thirddim}}}\times{({\K}^{\matrixsize{\firstdim}{\thirddim}})}^{*}&\rightarrow&{\K},\\
(\mat{M},\mat{N},\Transpose{\mat{O}})&\mapsto&\langle\mat{O},\MatrixProduct{M}{N}\rangle.
\end{array}
\end{equation}
The complete contraction of this tensor~\(\tensor{S}\) with an \emph{input} tensor~\({\mat{A}\tensorproduct\mat{B}\tensorproduct\mat{C}}\) yields its \emph{trilinear representation}:
\begin{align}\label{eq:StrassenThirdContraction}
&\mathcolor{\triadfive}{\left({a_{11}}+{a_{22}}\right)\tensorproduct\left({b_{11}}+{b_{22}}\right)\tensorproduct\left({c_{11}}+{c_{22}}\right)}\\
+&\mathcolor{\triadfour}{\left({a_{12}}-{a_{22}}\right)\tensorproduct\left({b_{21}}+{b_{22}}\right)\tensorproduct{c_{11}}}\\
+&\mathcolor{\triadseven}{\left(-{a_{11}}+{a_{21}}\right)\tensorproduct\left({b_{11}}+{b_{12}}\right)\tensorproduct{c_{22}}}\\
+&\mathcolor{\triadtwo}{\left({a_{11}}+{a_{12}}\right)\tensorproduct{b_{22}}\tensorproduct\left(-{c_{11}}+{c_{12}}\right)}\\
+&\mathcolor{\triadone}{{a_{11}}\tensorproduct\left({b_{12}}-{b_{22}}\right)\tensorproduct\left({c_{12}}+{c_{22}}\right)}\\
+&\mathcolor{\triadsix}{{a_{22}}\tensorproduct\left(-{b_{11}}+{b_{21}}\right)\tensorproduct\left({c_{11}}+{c_{21}}\right)}\\
+&\mathcolor{\triadthree}{\left({a_{21}}+{a_{22}}\right)\tensorproduct{b_{11}}\tensorproduct\left({c_{21}}-{c_{22}}\right)}.
\end{align}
This representation format is utilized to present our primary results in~\Cref{sec:TheAlgorithm}.
\subsection{Symmetries of matrix product tensor decomposition}\label{sec:Isotropies}
Remark that the matrix product is associated to~\({\Trace(\MatrixProduct{\mat{A}}{\MatrixProduct{\mat{B}}{\mat{C}}})}\) by~\Cref{eq:TrilinearForm}.
Furthermore, given three invertible matrices~\({\mat{U},\mat{V},\mat{W}}\) of suitable sizes and the classical trace properties, this trace is equal to:
\begin{equation}\label{eq:isotropy}
\begin{array}{l}
\Trace\bigl(\Transpose{(\MatrixProduct{\MatrixProduct{\mat{A}}{\mat{B}}}{\mat{C}})}\bigr)
=\Trace(\MatrixProduct{\mat{C}}{\MatrixProduct{\mat{A}}{\mat{B}}})
	=\Trace(\MatrixProduct{\mat{B}}{\MatrixProduct{\mat{C}}{\mat{A}}})\\[\smallskipamount]
\text{and to}\ \Trace\bigl(\MatrixProduct{\Inverse{\mat{U}}}{\MatrixProduct{\mat{A}}{\mat{V}}}
\cdot\Inverse{\mat{V}}\cdot{\mat{B}}\cdot{\mat{W}}\cdot\Inverse{\mat{W}}\cdot{\mat{C}}\cdot{\mat{U}}\bigr).
\end{array}
\end{equation}
These relations illustrate the next theorem and induce an isotropy action on matrix product tensor decompositions, presented below:
\begin{theorem}[{\cite[\S~2.8]{groot:1978a}}]\label{thm:groot}
The \emph{isotropy group} of the~\(\matrixsize{\firstdim}{\seconddim}\) by~\(\matrixsize{\seconddim}{\thirddim}\) matrix multiplication tensor is the semi-direct product
	\begin{equation}
	\SemiDirectProduct{\left(
		{{\emph{\textsc{psl}}^{\pm}\left({{\K}^{\firstdim}}\right)}}
	\times
	{{\emph{\textsc{psl}}^{\pm}\left({{\K}^{\seconddim}}\right)}}
	\times
	{{\emph{\textsc{psl}}^{\pm}\left({{\K}^{\thirddim}}\right)}}
	\right)}{\SymmetricGroup{3}},
	\end{equation}
where~\({\textsc{psl}^{\pm}}\) stands for the group of matrices of determinant~\({\pm{1}}\) and~\(\SymmetricGroup{3}\) for the symmetric group on~\(3\) elements.
\end{theorem}
As we do not harness the~\(\SymmetricGroup{3}\) part of this isotropy group, we do not give further details on its action.
Let us now describe how this group acts on tensor decompositions.
\begin{lemma}\label{lem:sandwiching}
Let~\(\tensor{T}\) denote a rank-one tensor~\({\mat{A}\tensorproduct\mat{B}\tensorproduct\mat{C}}\) and let~\(\Isotropy{g}\) denote the isotropy associated with matrices~\({(\mat{U}\times\mat{V}\times\mat{W})}\) in~\({{{\emph{\textsc{psl}}^{\pm}({{\K}^{\firstdim}})}}\times{{\emph{\textsc{psl}}^{\pm}({{\K}^{\seconddim}})}}\times{{\emph{\textsc{psl}}^{\pm}({{\K}^{\thirddim}})}}}\);
the action~\({\IsotropyAction{\Isotropy{g}}{\tensor{T}}}\) of~\(\Isotropy{g}\) on~\(\tensor{T}\) is the rank-one tensor:
\begin{equation}
{\left(\MatrixProduct{\InvTranspose{\mat{U}}}{\MatrixProduct{\mat{A}}{\Transpose{\mat{V}}}}\right)}
\tensorproduct
{\left(\MatrixProduct{\InvTranspose{\mat{V}}}{\MatrixProduct{\mat{B}}{\Transpose{\mat{W}}}}\right)}
\tensorproduct
{\left(\MatrixProduct{\InvTranspose{\mat{W}}}{\MatrixProduct{\mat{C}}{\Transpose{\mat{U}}}}\right)}.
\end{equation}
This action is extended by additivity to tensors with higher tensor rank.
Given two isotropies~\(\Isotropy{g}_{1}\) and~\(\Isotropy{g}_{2}\) defined respectively by the matrices~\({({\mat{U}_{1}},{\mat{V}_{1}},{\mat{W}_{1}})}\) and~\({({\mat{U}_{2}},{\mat{V}_{2}},{\mat{W}_{2}})}\) in~\({{{\emph{\textsc{psl}}^{\pm}({{\K}^{\firstdim}})}}\times{{\emph{\textsc{psl}}^{\pm}({{\K}^{\seconddim}})}}\times{{\emph{\textsc{psl}}^{\pm}({{\K}^{\thirddim}})}}}\), the composition~\({\Isotropy{g}_{1}\IsotropyComposition \Isotropy{g}_{2}}\) is given by~\({(\MatrixProduct{\mat{U}_{1}}{\mat{U}_{2}},{\MatrixProduct{\mat{V}_{1}}{\mat{V}_{2}}},{\MatrixProduct{\mat{W}_{1}}{\mat{W}_{2}}})}\) and we denote~\(\Isotropy{g}\IsotropyComposition\Isotropy{g}\) by~\(\Isotropy{g}^{2}\).
\end{lemma}
The following theorem recalls that any~\(\matrixsize{2}{2}\)-matrix multiplication algorithm with~\(7\) coefficient multiplications lies in a single orbit under the action of isotropies on Strassen tensor decomposition:
\begin{theorem}[{\cite[\S~0.1]{groot:1978}}]\label{thm:IsotropiesActTransitivelyOnOptimalAlgorithm}
The group~\({{\emph{\textsc{psl}}^{\pm}({{\K}^{\matrixsize{2}{2}}})}^{\times{3}}}\) \!acts transitively on the variety of fast algorithms multiplying~\(\matrixsize{2}{2}\)-matrices.
\end{theorem}
More generally, isotropy action on any decomposition may define another matrix product algorithm of the same tensor rank but with potentially more interesting features.
This fact is used in~\cite{jgd:2024:accurate,Dumas:2026:autoaccurate,Dumas:2026ac} to increase the numerical accuracy of some fast matrix multiplication algorithms.
There, we use the tensor decomposition orbit to express AlphaEvolve's~\(\FMMA{2}{2}{2}{7}\) and~\(\FMMA{3}{4}{7}{63}\) with more suitable coefficients.
We first illustrate the opposite in the following example.
\begin{example}
Consider the following isotropy:
\begin{equation}
\Isotropy{g}=
	\begin{bmatrix}
	1&0\\ 0&I\\
	\end{bmatrix}
	\times
	\begin{bmatrix}
	1&0\\ 0&1\\
	\end{bmatrix}
	\times
	\begin{bmatrix}
	1&0\\ 0&1\\
	\end{bmatrix}\quad\textup{where}\ {I=\sqrt{-1}}.
\end{equation}
The trilinear representation associated to the tensor decomposition~\({\IsotropyAction{\Isotropy{g}}{\tensor{S}}}\) is:
\begin{align}
&\left({a_{11}}-I{a_{22}}\right)\tensorproduct\left({b_{11}}+{b_{22}}\right)\tensorproduct\left({c_{11}}+I{c_{22}}\right)\\
+&\left({a_{12}}+I{a_{22}}\right)\tensorproduct\left({b_{21}}+{b_{22}}\right)\tensorproduct{c_{11}}\\
+&\left(-{a_{11}}-I{a_{21}}\right)\tensorproduct\left({b_{11}}+{b_{12}}\right)\tensorproduct{c_{22}}I\\
+&\left({a_{11}}+{a_{12}}\right)\tensorproduct{b_{22}}\tensorproduct\left(-{c_{11}}+{c_{12}}\right)\\
+&{a_{11}}\tensorproduct\left({b_{12}}-{b_{22}}\right)\tensorproduct\left({c_{12}}+I{c_{22}}\right)\\
-&I{a_{22}}\tensorproduct\left(-{b_{11}}+{b_{21}}\right)\tensorproduct\left({c_{11}}+I{c_{21}}\right)\\
+&\left(-I{a_{21}}-I{a_{22}}\right)\tensorproduct{b_{11}}\tensorproduct\left(I{c_{21}}-I{c_{22}}\right).
\end{align}
This last representation involves the complex number~\(I\) but it is clear that there exists at least one point in the orbit of Strassen's tensor decomposition involving only integers.
\end{example}
So one can consider the following question:
\begin{problem}
Is there a tensor decomposition in the orbit of~\(\FMMA{4}{4}{4}{48}\) introduced in~\cite{alphaevolve} that does not require complex numbers?
\end{problem}
In the following section, we describe now the experiments that allow us to find a rational point in the orbit of AlphaEvolve's~\(\FMMA{4}{4}{4}{48}\) complex tensor decomposition.
\section{Stabilizer of a tensor decomposition encoding the matrix product}\label{sec:Stabilizer}
\subsection{Invariant subgroups of the isotropy group}
Given a tensor decomposition, it can happen that a finite subgroup of the isotropy group described in~\Cref{thm:groot} sends this decomposition to itself acting as a permutation of its rank-one components.
This phenomenon occurs with several tensor decompositions such as
\begin{description}
	\item Strassen's~\(\FMMA{2}{2}{2}{7}\)~\cite{strassen:1969} with stabilizer~\(\SemiDirectProduct{\SymmetricGroup{3}}{\SymmetricGroup{3}}\),
	\item Hopcroft-Kerr's~\(\FMMA{2}{3}{3}{15}\)~\cite{hopcroft:1971} with~\(\SemiDirectProduct{\SymmetricGroup{3}}{\CyclicGroup{2}}\),
	\item Laderman's~\(\FMMA{3}{3}{3}{23}\)~\cite{laderman:1976a} with~\(\SemiDirectProduct{(\DirectProduct{\CyclicGroup{2}}{\CyclicGroup{2}})}{\SymmetricGroup{3}}\),
	\item Smirnov's~\(\FMMA{3}{3}{6}{40}\)~\cite{smirnov:2013a} with~\(\SemiDirectProduct{(\SemiDirectProduct{(\DirectProduct{\CyclicGroup{2}}{\CyclicGroup{2}})}{\SymmetricGroup{4}})}{\CyclicGroup{2}}\) and
	\item AlphaEvolve's~\(\FMMA{4}{4}{4}{48}\)~\cite{alphaevolve} with~\(\SemiDirectProduct{(\DirectProduct{\CyclicGroup{2}}{\CyclicGroup{4}})}{\DihedralGroup{4}}\)
\end{description}
(even if these tensor decompositions were found using different methods---other examples can be found at~\cite{Sedoglavic:FMMDB}).

\begin{remark}
Little is known about these stabilizers: their structural origins remain unclear, and aside from isolated cases (see~\cite{burichenko:2014}), it has not been established that the aforementioned groups constitute the full stabilizer group of their associated tensor decompositions.
\end{remark}
However, there are several theoretical and practical tools that could be used to find them.
First, given an encoding of a tensor decomposition, we could easily compute several of its invariants with respect to the isotropies' action.
For example, \Cref{def:RankAndType} and~\Cref{lem:sandwiching} imply directly the following results:
\begin{lemma}
The tensor decomposition rank and its type are invariants of the isotropies' action.
\end{lemma}
Using these invariants to prune an exhaustive search, together with an encoding of isotropies and their actions (see e.g.~\cite{Berger:2019aa}), one can design several computational tools to determine the stabilizer (or at least one of its subgroups).
Once computed, we can identify its associated abstract group using its Cayley table and thus, determine its presentation.
\par
Hence, the subgroup~\({\SemiDirectProduct{(\DirectProduct{\CyclicGroup{2}}{\DihedralGroup{4}})}{\CyclicGroup{2}}}\) of the known stabilizer of the~\(\FMMA{4}{4}{4}{48}\) tensor decomposition is given by the following generators:
\begin{align}
	\Isotropy{a}=& \begin{bmatrix} -I&0&0&0\\ 0&0&1&0\\ 0&1&0&0\\ 0&0&0&I\end{bmatrix}&\times&\begin{bmatrix} 0&0&1&0\\ 0&-1&0&0\\ -1&0&0&0\\ 0&0&0&1\end{bmatrix}&\times&\begin{bmatrix} -1&0&0&0\\ 0&1&0&0\\ 0&1&0&-1\\ 1&0&1&0\end{bmatrix}, \label{eq:abcd:a}\\
	\Isotropy{b}=&\begin{bmatrix} 0&0&0&1\\ 0&0&1&0\\ 0&1&0&0\\ 1&0&0&0\end{bmatrix}&\times&\begin{bmatrix} 1&0&0&0\\ 0&1&0&0\\ 0&0&-1&0\\ 0&0&0&1\end{bmatrix}&\times&\begin{bmatrix} 1&0&0&0\\ 0&1&0&0\\ -1&1&-1&0\\ 0&0&0&1\end{bmatrix}, \label{eq:abcd:b}\\
	\Isotropy{c}=&\begin{bmatrix} 0&0&0&1\\ 0&1&0&0\\ 0&0&-1&0\\ -1&0&0&0\end{bmatrix}&\times&\begin{bmatrix} 1&0&0&0\\ 0&0&0&I\\ 0&0&-1&0\\ 0&I&0&0\end{bmatrix}&\times&\begin{bmatrix} 0&-1&0&0\\ 1&0&0&0\\ 0&1&-1&0\\ 1&0&0&1\end{bmatrix}, \label{eq:abcd:c}\\
	\Isotropy{d}=&\begin{bmatrix} 0&0&0&1\\ 0&0&-1&0\\ 0&1&0&0\\ -1&0&0&0\end{bmatrix}&\times&\begin{bmatrix} 1&0&0&0\\ 0&1&0&0\\ 0&0&1&0\\ 0&0&0&-1\end{bmatrix}&\times&\begin{bmatrix} 0&1&0&0\\ 1&0&0&0\\ 1&-1&1&0\\ 0&0&0&-1\end{bmatrix}\label{eq:abcd:d}
\end{align}
and the following relations:
\begin{multline}\label{eq:PresentationRelations}
{\Isotropy{a}^{4}=\Isotropy{b}^{2}=\Isotropy{d}^{2}=1},\quad{\Isotropy{c}^{2}=\Isotropy{a}^{2}},\quad{\Isotropy{b}\IsotropyComposition\Isotropy{a}\IsotropyComposition\Isotropy{b}\IsotropyComposition\Isotropy{a}=1},\quad{\Isotropy{a}\IsotropyComposition\Isotropy{c}=\Isotropy{c}\IsotropyComposition\Isotropy{a}},\\
	{\Isotropy{a}\IsotropyComposition\Isotropy{d}=\Isotropy{d}\IsotropyComposition\Isotropy{a}},\quad{\Isotropy{b}\IsotropyComposition\Isotropy{c}=\Isotropy{c}\IsotropyComposition\Isotropy{b}},\quad{\Isotropy{b}\IsotropyComposition\Isotropy{d}=\Isotropy{d}\IsotropyComposition\Isotropy{b}},\quad{\Isotropy{d}\IsotropyComposition\Isotropy{c}\IsotropyComposition\Isotropy{d}=\Isotropy{c}^{3}}.
\end{multline}
The generators~(\Cref{eq:abcd:a,eq:abcd:b,eq:abcd:c,eq:abcd:d}) were chosen arbitrarily by inspection to fit the above constraints.
\subsection{The stabilizer as a blueprint of the associated tensor decomposition}
Each element of the stabilizer acts on the tensor decomposition by permuting its rank-one summands.
The orbits of such actions were used in~\cite{Grochow:2016aa} to propose a construction of Strassen's tensor decomposition~\(\FMMA{2}{2}{2}{7}\) and to construct a family of tensor decompositions~\(\FMMA{n}{n}{n}{n^{3}-n+1}\) based on~\(\SymmetricGroup{n}\) for~\(n\) greater than~\(2\).
Symmetries were also used in~\cite{Sedoglavic:2017aa} to show how to deduce Laderman's~\(\FMMA{3}{3}{3}{23}\) from Winograd's variant of~\(\FMMA{2}{2}{2}{7}\) and in~\cite{Moosbauer:2025aa} to construct a tensor decomposition~\(\FMMA{5}{5}{5}{93}\) with stabilizer~\(\CyclicGroup{3}\).
\par
The type of the~\(\FMMA{4}{4}{4}{48}\) tensor decomposition from~\cite{alphaevolve} is~\(16{X^{2}Y^{2}Z^{2}+32XYZ}\).
Furthermore, it is composed of two orbits: there are~\(16\) rank-one tensors of type~\({[2,2,2]}\) whose orbits under the actions of the stabilizer~\({\SemiDirectProduct{(\DirectProduct{\CyclicGroup{2}}{\DihedralGroup{4}})}{\CyclicGroup{2}}}\) are of order~\(16\) and~\(32\) rank-one tensors of type~\({[1,1,1]}\) whose orbits under the actions of this stabilizer are of order~\(32\).
\par
Hence, there are two \emph{seeds} (say):
\begin{align}
\tau_{4}=& \begin{bmatrix} I&1&-1&1\\ -I&1&-1&-1\\ I&1&-1&1\\ 1&I&-I&-I\end{bmatrix} \tensorproduct \begin{bmatrix} I&0&-I&0\\ 0&I&I&0\\ 0&-I&-I&0\\ -1&0&1&0\end{bmatrix} \tensorproduct \begin{bmatrix} 0&0&0&0\\ -1-I&1-I&-1-I&-1-I\\ 1&-1&1&I\\ I&I&I&1\end{bmatrix}/8,\label{eq:tau:4}\\
 \tau_{1} =& \begin{bmatrix} 1+I&1+I&0&0\\ 1-I&1-I&0&0\\ 1-I&1-I&0&0\\ 1-I&1-I&0&0\end{bmatrix}\tensorproduct\begin{bmatrix} -1&0&0&0\\ -1&0&0&0\\ 1&0&0&0\\ -I&0&0&0\end{bmatrix}\tensorproduct\begin{bmatrix} I&-1&-I&-I\\ -I&1&I&I\\ 0&0&0&0\\ I&-1&-I&-I\end{bmatrix}/8\label{eq:tau:1}
\end{align}
such that this tensor decomposition could be written as follows:
\begin{equation}\label{eq:GroupAverage}
\FMMA{4}{4}{4}{48}={\frac{1}{2}\sum_{\Isotropy{g}\in{\SemiDirectProduct{(\DirectProduct{\CyclicGroup{2}}{\DihedralGroup{4}})}{\CyclicGroup{2}}}}\IsotropyAction{\Isotropy{g}}{\tau_{4}}}+{\sum_{\Isotropy{g}\in{\SemiDirectProduct{(\DirectProduct{\CyclicGroup{2}}{\DihedralGroup{4}})}{\CyclicGroup{2}}}}\IsotropyAction{\Isotropy{g}}{\tau_{1}}}.
\end{equation}
Any isotropy~\(\rho\) acting on the tensor decomposition acts on~\Cref{eq:GroupAverage} as follows:
\begin{equation}\label{eq:ActionOnAverageByGroup}
{\IsotropyAction{\rho}{\FMMA{4}{4}{4}{48}}}
=\sum_{\Isotropy{g}\in{\SemiDirectProduct{(\DirectProduct{\CyclicGroup{2}}{\DihedralGroup{4}})}{\CyclicGroup{2}}}}
\IsotropyAction{({\rho}\IsotropyComposition{\Isotropy{g}}\IsotropyComposition{\rho^{-1}})}{\frac{(\IsotropyAction{\rho}{\tau_{4}})}{2}}
+\IsotropyAction{({\rho}\IsotropyComposition{\Isotropy{g}}\IsotropyComposition{\rho^{-1}})}{(\IsotropyAction{\rho}{\tau_{1}})}.
\end{equation}
Consequently, a sufficient condition for an isotropy~\(\rho\) to produce a tensor \(\IsotropyAction{\rho}{\FMMA{4}{4}{4}{48}}\) over the rationals, is that it does so on the two seeds (i.e.\ ensuring that~\(\IsotropyAction{\rho}{\tau_4}\) and~\(\IsotropyAction{\rho}{\tau_1}\)are rationals) and sends, by similarity transformation, each isotropy~\(\Isotropy{g}\) of the stabilizer's subgroup~\({\SemiDirectProduct{(\DirectProduct{\CyclicGroup{2}}{\DihedralGroup{4}})}{\CyclicGroup{2}}}\) to a rational one.
Using the presentation of the stabilizer's subgroup~\({\SemiDirectProduct{(\DirectProduct{\CyclicGroup{2}}{\DihedralGroup{4}})}{\CyclicGroup{2}}}\), the second condition only needs to be verified on its four generators~\({\Isotropy{a},\Isotropy{b},\Isotropy{c},\Isotropy{d}}\).
\begin{lemma}\label{lem:MagicIsotropy}
The following isotropy
	\begin{equation}\label{eq:MagicIsotropy}
	\Isotropy{p}=
	{\begin{bmatrix}
		I&0&0&I\\
		0&1&I&0\\
		0&-I&-1&0\\
		-1&0&0&1
	\end{bmatrix}}
	\times
	{\begin{bmatrix}
		I&0&0&1\\
		0&-I&-I&0\\
		0&-I&I&0\\
		-I&0&0&1
	\end{bmatrix}}
	\times
	{\begin{bmatrix}
		1&0&0&0\\
		0&1&0&0\\
		0&0&1&0\\
		0&0&0&1
	\end{bmatrix}},
\end{equation}
ensures that the following four isotropies and two rank-one tensors:
	\begin{equation}
		{\Isotropy{p}}\IsotropyComposition{\Isotropy{g}}\IsotropyComposition{\Isotropy{p}^{-1}} \ \textup{with}\ \Isotropy{g}\in\{ \Isotropy{a}, \Isotropy{b}, \Isotropy{c}, \Isotropy{d} \},\quad \IsotropyAction{\Isotropy{p}}{\InvTranspose{\tau_{i}}}\ \textup{with}\ i\in\{1,4\}
	\end{equation}
	obtained by~\(\Isotropy{p}\) and~\Cref{eq:abcd:a,eq:abcd:b,eq:abcd:c,eq:abcd:d,eq:tau:4,eq:tau:1} are defined over the rational field.
\end{lemma}
This lemma implies that there is a rational point in AlphaEvolve's~\(\FMMA{4}{4}{4}{48}\) orbit.
We now present our main result in the following section.
\section{A rational point in the orbit of AlphaEvolve's \texorpdfstring{\(\FMMA{4}{4}{4}{48}\)}{<4x4x4:48>} tensor decomposition}\label{sec:TheAlgorithm}
The isotropy~\(\Isotropy{p}\) introduced in~\Cref{eq:MagicIsotropy} sends the original AlphaEvolve's~\(\FMMA{4}{4}{4}{48}\) complex-valued tensor decomposition to the following rational one, up to some inner shifts or change of signs\footnote{Rank-one tensors are defined in a projective space:~\({{{(\sum{a_{ij}})}\tensorproduct{(\sum{b_{ij}})}\tensorproduct{(\sum{c_{ij}})}}={{(\alpha\sum{a_{ij}})}\tensorproduct{(\beta\sum{b_{ij}})}\tensorproduct{(\gamma\sum{c_{ij}})}}}\) when~\({\alpha\beta\gamma=1}\).}, as shown in~\Crefrange{eq:firstprod}{eq:lastprod}.
\begin{align}
	\label{eq:firstprod}
m_{1}=&\begin{aligned}[t]
	\left(\begin{array}{c}
		-{a_{11}}-{a_{21}}+{a_{31}}-{a_{41}}
		+{a_{12}}+{a_{22}}-{a_{32}}+{a_{42}}\\
		+{a_{13}}+{a_{23}}-{a_{33}}+{a_{43}}
		+{a_{14}}+{a_{24}}-{a_{34}}+{a_{44}}
	\end{array}\right)\tensorproduct
	\left({b_{31}}+{b_{41}}\right)\tensorproduct
	\frac{\left(\begin{array}{c}
		{c_{41}} -{c_{31}}
		+{c_{32}} \\ -{c_{42}}
		+{c_{44}} -{c_{34}}
		\end{array}\right)}{4},
	\end{aligned}\\
m_{2}=&\begin{aligned}[t]
	\left( {a_{21}} -{a_{11}} +{a_{31}}-{a_{41}}\right)\tensorproduct
	\left({b_{12}}-{b_{22}}+{b_{14}}-{b_{24}}\right)\tensorproduct
	\lfrac{\left({c_{21}}+{c_{24}}-{c_{11}}-{c_{14}}\right)}{2},
	\end{aligned}\\
m_{3}=&\begin{aligned}[t]
	\left({-a_{13}}+{a_{23}}+{a_{33}}+{a_{43}}\right)\tensorproduct
	\left({b_{12}}+{b_{32}}\right)\tensorproduct
	\lfrac{\left({c_{22}}-{c_{23}}+{c_{42}}-{c_{43}}\right)}{2},
	\end{aligned}\\
m_{4}=&\begin{aligned}[t]
	\left(\begin{array}{c}
		-{a_{13}}-{a_{23}}+{a_{33}}+{a_{43}}\\
		-{a_{14}}-{a_{24}}-{a_{34}}-{a_{44}}
		\end{array}\right)\tensorproduct
	\left(\begin{array}{c}{b_{41}}+{b_{32}}\\ +{b_{33}}-{b_{43}}\end{array}\right)\tensorproduct
	\lfrac{\left(\begin{array}{c}
		{c_{13}} -({c_{12}}+{c_{22}})/2 +{c_{14}} +{c_{23}}\\
		+{c_{24}}+{c_{33}}-{c_{34}}+{c_{43}}-{c_{44}}
		\end{array}\right)}{4},
	\end{aligned}\\
m_{5}=&\begin{aligned}[t]
	\left(\begin{array}{c}
	{a_{11}}-{a_{21}}+{a_{31}}+{a_{41}}
	+{a_{12}}-{a_{22}}+{a_{32}}+{a_{42}}\\
	+{a_{13}}-{a_{23}}+{a_{33}}+{a_{43}}
	-{a_{14}}+{a_{24}}-{a_{34}}-{a_{44}}
	\end{array}\right)\tensorproduct
	\left(\begin{array}{c}
		-{b_{11}}-{b_{21}}
		+{b_{13}} \\ +{b_{23}}
		+{b_{14}}+{b_{24}}
		\end{array}
	\right)\tensorproduct
	\frac{\left({c_{34}}+{c_{44}}\right)}{4},
	\end{aligned}\\
m_{6}=&\begin{aligned}[t]
	\left(\begin{array}{c}
	{a_{11}}-{a_{21}}-{a_{31}}+{a_{41}}\\
	+{a_{12}}-{a_{22}}-{a_{32}}+{a_{42}}\\
	-{a_{13}}+{a_{23}}+{a_{33}}-{a_{43}}\\
	+{a_{14}}-{a_{24}}-{a_{34}}+{a_{44}}
	\end{array}\right)\tensorproduct
	\left({%
		{b_{22}}+{b_{42}}%
		+{b_{24}}+{b_{44}}
	}\right)\tensorproduct
	\frac{\left(\begin{array}{c}
		{c_{21}}-{c_{22}}+{c_{23}}\\
		-{c_{41}}+{c_{42}}-{c_{43}}
	\end{array}\right)}{4},
	\end{aligned}\\
m_{7}=&\begin{aligned}[t]
	\left(-{a_{14}}+{a_{24}}+{a_{34}}+{a_{44}}\right)\tensorproduct
	\left( {b_{22}} -{b_{42}} \right)\tensorproduct
	\lfrac{\left({c_{12}}-{c_{14}}-{c_{32}}+{c_{34}}\right)}{2},
	\end{aligned}\\
m_{8}=&\begin{aligned}[t]
	\left(\begin{array}{c}
	 {a_{11}}-{a_{21}}+{a_{31}}-{a_{41}}+{a_{12}}-{a_{22}}+{a_{32}}-{a_{42}}\\
	+{a_{13}}-{a_{23}}+{a_{33}}-{a_{43}}-{a_{14}}+{a_{24}}-{a_{34}}+{a_{44}}
	\end{array}\right)\tensorproduct
	\left(\begin{array}{c}
		{b_{11}} -{b_{14}}\\
		+{b_{31}} -{b_{34}}
	\end{array}\right)\tensorproduct
	\frac{\left({c_{24}}+{c_{44}}\right)}{4},
	\end{aligned} \\
m_{9}=&\begin{aligned}[t]
	\left({\begin{array}{c}
		-{a_{31}}-{a_{41}}\\
		+{a_{32}}+{a_{42}}\\
		-{a_{13}}+{a_{23}}\\
		-{a_{14}}+{a_{24}}
	\end{array}}\right)\tensorproduct
	\left(\begin{array}{c}
	{b_{11}}-{b_{21}}-{b_{31}}-{b_{41}}\\
	+{b_{12}}-{b_{22}}+{b_{32}}+{b_{42}}\\
	-{b_{13}}+{b_{23}}+{b_{33}}+{b_{43}}\\
	-{b_{14}}+{b_{24}}+{b_{34}}+{b_{44}}
	\end{array}\right)\tensorproduct
	{\lfrac{\left(\begin{array}{c}
	{c_{11}}-{c_{12}}-{c_{13}}+{c_{14}}\\
		-{c_{21}}+{c_{22}}+{c_{23}}-{c_{24}}\\
	+{c_{31}}-{c_{32}}+{c_{33}}+{c_{34}}\\
		+{c_{41}}-{c_{42}}+{c_{43}}+{c_{44}}
	\end{array}\right)}{8}},
	\end{aligned}\\
m_{10}=&\begin{aligned}[t]
	\left({a_{11}}+{a_{21}}+{a_{31}}-{a_{41}}\right)\tensorproduct
	\left({b_{12}}-{b_{32}}+{b_{13}}-{b_{33}}+{b_{14}}-{b_{34}}\right)\tensorproduct
	\lfrac{\left({c_{11}}+{c_{14}}+{c_{31}}+{c_{34}}\right)}{2},
	\end{aligned}\\
m_{11}=&\begin{aligned}[t]
	\left(\begin{array}{c}
	-{a_{11}}+{a_{21}}+{a_{31}}-{a_{41}}-{a_{12}}+{a_{22}}+{a_{32}}-{a_{42}}\\
	-{a_{13}}+{a_{23}}+{a_{33}}-{a_{43}}+{a_{14}}-{a_{24}}-{a_{34}}+{a_{44}}
	\end{array}\right)\tensorproduct
	\left(\begin{array}{c}
		{b_{22}}-{b_{42}}\\
		+{b_{24}}-{b_{44}}
	\end{array}\right)\tensorproduct
	\frac{\left(-{c_{14}}+{c_{34}}\right)}{4},
	\end{aligned}\\
m_{12}=&\begin{aligned}[t]
	\left(\begin{array}{c}
		-{a_{21}}-{a_{41}}\\
		+{a_{12}}+{a_{32}}\\
		+{a_{23}}+{a_{43}}\\
		+{a_{14}}+{a_{34}}
	\end{array}\right)\tensorproduct
	\left(\begin{array}{c}
	-{b_{11}}+{b_{21}}+{b_{31}}+{b_{41}}\\
	-{b_{12}}-{b_{22}}+{b_{32}}-{b_{42}}\\
	-{b_{13}}-{b_{23}}+{b_{33}}-{b_{43}}\\
	+{b_{14}}-{b_{24}}-{b_{34}}-{b_{44}}
	\end{array}\right)\tensorproduct
	\lfrac{\left(\begin{array}{c}
	2({c_{11}}-{c_{12}})+{c_{13}}\\
		+{c_{14}}+{c_{23}}-{c_{24}} \\
	+2({c_{31}}-{c_{32}})+{c_{33}}\\
		+{c_{34}}+{c_{43}}-{c_{44}}
	\end{array}\right)}{8},
	\end{aligned}\\
m_{13}=&\begin{aligned}[t]
	\left({a_{13}}-{a_{23}}+{a_{33}}+{a_{43}}\right)\tensorproduct
	\left({b_{11}}-{b_{31}}-{b_{13}}+{b_{33}}-{b_{14}}+{b_{34}}\right)\tensorproduct
	\lfrac{\left(-{c_{12}}+{c_{13}}-{c_{32}}+{c_{33}}\right)}{2},
	\end{aligned}\\
m_{14}=&\begin{aligned}[t]
\left({a_{13}}+{a_{23}}-{a_{33}}-{a_{43}}\right)\tensorproduct
\left({b_{31}}-{b_{41}}-{b_{33}}+{b_{43}}\right)\tensorproduct
	\lfrac{\left({c_{12}}-{c_{13}}+{c_{22}}-{c_{23}}\right)}{2},
	\end{aligned}\\
m_{15}=&\begin{aligned}[t]
	\left(\begin{array}{c}
	{a_{11}}+{a_{21}}\\
	-{a_{31}}+{a_{41}}\\
	+{a_{13}}-{a_{23}}\\
	-{a_{33}}-{a_{43}}
	\end{array}\right)\tensorproduct
	\left(-{b_{31}}+{b_{12}}\right)\tensorproduct
	\lfrac{\left(\begin{array}{c}
	-{c_{11}}+{c_{12}}-{c_{13}}-{c_{14}}
	+{c_{21}}+{c_{22}}-{c_{23}}+{c_{24}}\\
	 {c_{31}}-{c_{32}}+{c_{33}}+{c_{34}}
	+{c_{41}}+{c_{42}}-{c_{43}}+{c_{44}}
	\end{array}\right)}{4},
	\end{aligned}\\
m_{16}=&\begin{aligned}[t]
	\left(\begin{array}{c}
		{a_{11}}-{a_{21}}+{a_{31}}-{a_{41}}\\
		+{a_{12}}-{a_{22}}+{a_{32}}-{a_{42}}\\
		-{a_{13}}+{a_{23}}-{a_{33}}+{a_{43}}\\
		+{a_{14}}-{a_{24}}+{a_{34}}-{a_{44}}
	\end{array}\right)\tensorproduct
	\left(\begin{array}{c}
		{b_{11}} -{b_{14}} \\
		-{b_{31}} +{b_{34}}
		\end{array}\right)\tensorproduct
	\frac{\left({c_{11}}-{c_{12}}+{c_{13}}+{c_{31}}-{c_{32}}+{c_{33}}\right)}{4}
	\end{aligned}\\
m_{17}=&\begin{aligned}[t]
	\left(\begin{array}{c}
		{a_{11}}+{a_{21}}-{a_{31}}-{a_{41}}\\
		+{a_{12}}+{a_{22}}+{a_{32}}+{a_{42}}
	\end{array}\right)\tensorproduct
	\left(\begin{array}{c}
		{b_{21}}-{b_{12}}\\
		-{b_{13}}-{b_{23}}
	\end{array}\right)\tensorproduct
	\lfrac{\left(\begin{array}{c}
		2({c_{31}}+{c_{41}}) -{c_{13}} -{c_{23}} +{c_{33}} \\
		+{c_{43}} +{c_{14}} +{c_{24}} +{c_{34}}+{c_{44}}
	\end{array}\right)}{4},
	\end{aligned}\\
m_{18}=&\begin{aligned}[t]
	\left(\begin{array}{c}
	-{a_{11}}-{a_{21}}+{a_{31}}+{a_{41}}\\
		+{a_{12}}+{a_{22}}-{a_{32}}-{a_{42}}\\
	+{a_{13}}+{a_{23}}-{a_{33}}-{a_{43}}\\
		+{a_{14}}+{a_{24}}-{a_{34}}-{a_{44}}
	\end{array}\right)\tensorproduct
	\left(\begin{array}{c}
		{b_{21}}+{b_{41}}\\
		-{b_{23}}-{b_{43}}
	\end{array}\right)\tensorproduct
	\frac{\left(\begin{array}{c}
		{c_{21}}-{c_{22}}+{c_{24}}-{c_{41}}+{c_{42}}-{c_{44}}
		\end{array}\right)}{4},
	\end{aligned}\\
m_{19}=&\begin{aligned}[t]
	\left(\begin{array}{c}
		-{a_{21}}+{a_{41}}\\
		+{a_{12}}-{a_{32}}\\
		+{a_{23}}-{a_{43}}\\
		+{a_{14}}-{a_{34}}
	\end{array}\right)\tensorproduct
	\left(\begin{array}{c}
	{b_{11}}+{b_{21}}-{b_{31}}+{b_{41}}\\
	-{b_{12}}+{b_{22}}+{b_{32}}+{b_{42}}\\
	-{b_{13}}-{b_{23}}+{b_{33}}-{b_{43}}\\
	-{b_{14}}+{b_{24}}+{b_{34}}+{b_{44}}
	\end{array}\right)\tensorproduct
	\lfrac{\left(\begin{array}{c}
	2({c_{41}} -{c_{21}}) +2({c_{22}} -{c_{42}}) \\
	-{c_{13}} -{c_{23}} +{c_{33}} +{c_{43}} \\
	+{c_{14}} -{c_{24}} -{c_{34}} +{c_{44}} \\
	\end{array}\right)}{8},
	\end{aligned}
\end{align} \begin{align}
m_{20}=&\begin{aligned}[t]
	\left(\begin{array}{c}
	-{a_{11}}-{a_{21}}+{a_{31}}+{a_{41}}+{a_{12}}+{a_{22}}-{a_{32}}-{a_{42}}\\
	-{a_{13}}-{a_{23}}+{a_{33}}+{a_{43}}-{a_{14}}-{a_{24}}+{a_{34}}+{a_{44}}
	\end{array}\right)\tensorproduct
	\left(\begin{array}{c}
		-{b_{21}}+{b_{41}}\\
		+{b_{23}}-{b_{43}}
	\end{array}\right)\tensorproduct
	\frac{\left({c_{13}}-{c_{33}}\right)}{4},
	\end{aligned}\\
m_{21}=&\begin{aligned}[t]
	\left({a_{13}}-{a_{23}}-{a_{33}}+{a_{43}}\right)\tensorproduct
	\left({b_{32}}+{b_{42}}+{b_{34}}+{b_{44}}\right)\tensorproduct
	\lfrac{\left({c_{42}}-{c_{32}}+{c_{33}}-{c_{43}}\right)}{2},
	\end{aligned}\\
m_{22}=&\begin{aligned}[t]
	\left(\begin{array}{c}
	{a_{11}}+{a_{31}}-{a_{22}}-{a_{42}}\\
	+{a_{13}} +{a_{33}}+{a_{24}}+{a_{44}}
	\end{array}\right)\tensorproduct
	\left(\begin{array}{c}
	{b_{11}}-{b_{21}}+{b_{31}}+{b_{41}}\\
	+{b_{12}}+{b_{22}}+{b_{32}}-{b_{42}}\\
	+{b_{13}}+{b_{23}}+{b_{33}}-{b_{43}}\\
	-{b_{14}}+{b_{24}}-{b_{34}}-{b_{44}}
	\end{array}\right)\tensorproduct
	\frac{\left(\begin{array}{c}
		{c_{13}}-{c_{14}}-{c_{23}}-{c_{24}}\\
		+{c_{33}} -{c_{34}}-{c_{43}}-{c_{44}}
	\end{array}\right)}{8},
	\end{aligned}\\
m_{23}=&\begin{aligned}[t]
	\left({a_{14}}-{a_{24}}+{a_{34}}+{a_{44}}\right)\tensorproduct
	\left(\begin{array}{c}
		-{b_{21}}-{b_{41}}
		+{b_{23}}+{b_{43}}
		+{b_{24}}+{b_{44}}
	\end{array}\right)\tensorproduct
	\lfrac{\left({c_{22}}-{c_{24}}-{c_{42}}+{c_{44}}\right)}{2},
	\end{aligned}\\
m_{24}=&\begin{aligned}[t]
	\left(\begin{array}{c}
	{a_{11}}-{a_{21}}+{a_{31}}+{a_{41}}-{a_{12}}+{a_{22}}-{a_{32}}-{a_{42}}\\
	+{a_{13}}-{a_{23}}+{a_{33}}+{a_{43}}+{a_{14}}-{a_{24}}+{a_{34}}+{a_{44}}
	\end{array}\right)\tensorproduct
	\left(\begin{array}{c}
		{b_{11}}-{b_{21}}-{b_{13}}\\
		+{b_{23}}-{b_{14}}+{b_{24}}
	\end{array}\right)\tensorproduct
	\frac{\left( {c_{23} -{c_{13}} }\right)}{4},\\
	\end{aligned}\\
m_{25}=&\begin{aligned}[t]
	\left(\begin{array}{c}
		{a_{11}}-{a_{21}}\\
		+{a_{12}}-{a_{22}}\\
		+{a_{33}} -{a_{34}}\\
		+{a_{43}} -{a_{44}}
	\end{array}\right)\tensorproduct
	\left(\begin{array}{c}
	{b_{11}}+{b_{21}}-{b_{31}}+{b_{41}}\\
	+{b_{12}}+{b_{22}}+{b_{32}}-{b_{42}}\\
	-{b_{13}}-{b_{23}}+{b_{33}}-{b_{43}}\\
	-{b_{14}}-{b_{24}}+{b_{34}}-{b_{44}}
	\end{array}\right)\tensorproduct
	\lfrac{\left(\begin{array}{c}
	-{c_{11}}+{c_{12}}-{c_{13}}-{c_{14}}\\
	+{c_{21}}-{c_{22}}+{c_{23}}+{c_{24}}\\
	-{c_{31}}+{c_{32}}-{c_{33}}+{c_{34}}\\
	-{c_{41}}+{c_{42}}-{c_{43}}+{c_{44}}
	\end{array}\right)}{8},
	\end{aligned}\\
m_{26}=&\begin{aligned}[t]
	\left(\begin{array}{c}
	{a_{11}}+{a_{21}}+{a_{31}}-{a_{41}}-{a_{12}}-{a_{22}}-{a_{32}}+{a_{42}}\\
	+{a_{13}}+{a_{23}}+{a_{33}}-{a_{43}}+{a_{14}}+{a_{24}}+{a_{34}}-{a_{44}}
	\end{array}\right)\tensorproduct
	\left(\begin{array}{c}
		{b_{32}}+{b_{42}}+{b_{33}}\\
		+{b_{43}}+{b_{34}}+{b_{44}}
	\end{array}\right)\tensorproduct
	\frac{\left( {c_{33}} -{c_{43}} \right)}{4},
	\end{aligned}\\
m_{27}=&\begin{aligned}[t]
	\left({a_{12}}+{a_{22}}+{a_{32}}+{a_{42}}\right)\tensorproduct
	\left({b_{12}}+{b_{22}}+{b_{13}}+{b_{23}}\right)\tensorproduct
	\lfrac{\left({c_{31}}+{c_{33}}+{c_{41}}+{c_{43}}\right)}{2},
	\end{aligned}\\
m_{28}=&\begin{aligned}[t]
	\left(\begin{array}{c}
	{a_{31}}-{a_{41}}\\
	-{a_{32}}+{a_{42}}\\
	+{a_{13}}+{a_{23}}\\
	+{a_{14}}+{a_{24}}
	\end{array}\right)\tensorproduct
	\left(\begin{array}{c}
	{b_{11}}-{b_{21}}+{b_{31}}+{b_{41}}\\
	-{b_{12}}+{b_{22}}+{b_{32}}+{b_{42}}\\
	-{b_{13}}+{b_{23}}+{b_{33}}+{b_{43}}\\
	-{b_{14}}+{b_{24}}+{b_{34}}+{b_{44}}
	\end{array}\right)\tensorproduct
	\lfrac{\left(\begin{array}{c}
	{c_{11}}-{c_{12}}+{c_{13}}+{c_{14}}\\
	+{c_{21}}-{c_{22}}+{c_{23}}+{c_{24}}\\
	+{c_{31}}-{c_{32}}-{c_{33}}+{c_{34}}\\
	-{c_{41}}+{c_{42}}+{c_{43}}-{c_{44}}
	\end{array}\right)}{8},
	\end{aligned}\\
m_{29}=&\begin{aligned}[t]
	\left(\begin{array}{c}
	{a_{11}}-{a_{21}}-{a_{31}}-{a_{41}}-{a_{12}}+{a_{22}}+{a_{32}}+{a_{42}}\\
	-{a_{13}}+{a_{23}}+{a_{33}}+{a_{43}}-{a_{14}}+{a_{24}}+{a_{34}}+{a_{44}}
	\end{array}\right)\tensorproduct
	\left( {b_{22}} -{b_{12}} \right)\tensorproduct
	\frac{\left(\begin{array}{c}
		{c_{11}}-{c_{12}}+{c_{14}}\\
		-{c_{21}}+{c_{22}}-{c_{24}}
	\end{array}\right)}{4},
	\end{aligned}\\
m_{30}=&\begin{aligned}[t]
	\left({a_{14}}+{a_{24}}+{a_{34}}+{a_{44}}\right)\tensorproduct
	\left({b_{42}}+{b_{43}}-{b_{32}}-{b_{33}}\right)\tensorproduct
	\lfrac{\left({c_{12}}+{c_{22}}-{c_{14}}-{c_{24}}\right)}{2},
	\end{aligned}\\
m_{31}=&\begin{aligned}[t]
	\left({a_{12}}-{a_{22}}+{a_{32}}-{a_{42}}\right)\tensorproduct
	\left({b_{11}}-{b_{21}}-{b_{14}}+{b_{24}}\right)\tensorproduct
	\lfrac{\left({c_{21}}-{c_{11}}+{c_{23}}-{c_{13}}\right)}{2},
	\end{aligned}\\
m_{32}=&\begin{aligned}[t]
	\left(\begin{array}{c}
	{a_{11}}-{a_{31}}\\
	-{a_{22}}+{a_{42}}\\
	+{a_{13}}-{a_{33}}\\
	+{a_{24}}-{a_{44}}
	\end{array}\right)\tensorproduct
	\left(\begin{array}{c}
	{b_{11}}+{b_{21}}+{b_{31}}-{b_{41}}\\
	-{b_{12}}+{b_{22}}-{b_{32}}-{b_{42}}\\
	-{b_{13}}-{b_{23}}-{b_{33}}+{b_{43}}\\
	-{b_{14}}+{b_{24}}-{b_{34}}-{b_{44}}
	\end{array}\right)\tensorproduct
	\frac{\left(\begin{array}{c}
		-{c_{13}}-{c_{14}}+{c_{23}}-{c_{24}}\\
		+{c_{33}}+{c_{34}}-{c_{43}}+{c_{44}}
	\end{array}\right)}{8},
	\end{aligned}\\
m_{33}=&\begin{aligned}[t]
	\left(\begin{array}{c}
	-{a_{11}}+{a_{21}}+{a_{31}}+{a_{41}}\\
	-{a_{12}}+{a_{22}}+{a_{32}}+{a_{42}}\\
	+{a_{13}}-{a_{23}}-{a_{33}}-{a_{43}}\\
	-{a_{14}}+{a_{24}}+{a_{34}}+{a_{44}}
	\end{array}\right)\tensorproduct
	\left(-{b_{12}}-{b_{22}}\right)\tensorproduct
	\frac{\left(\begin{array}{c}
		{c_{31}}-{c_{32}}+{c_{33}}\\
		+{c_{41}}-{c_{42}}+{c_{43}}
	\end{array}\right)}{4},
	\end{aligned}\\
m_{34}=&\begin{aligned}[t]
	\left(\begin{array}{c}
	{-a_{13}}+{a_{23}}+{a_{33}}-{a_{43}}\\
	-{a_{14}}+{a_{24}}-{a_{34}}+{a_{44}}
	\end{array}\right)\tensorproduct
	\left(\begin{array}{c}
	-{b_{31}}+{b_{42}}\\
	+{b_{34}}+{b_{44}}
	\end{array}\right)\tensorproduct
	\lfrac{\left(\begin{array}{c}
	2({c_{42}}-{c_{32}}) +{c_{13}} -{c_{23}} +{c_{33}} \\
	-{c_{43}} -{c_{14}} +{c_{24}} +{c_{34}} -{c_{44}}
	\end{array}\right)}{4},
	\end{aligned}\\
m_{35}=&\begin{aligned}[t]
	\left(\begin{array}{c}
	{a_{11}}+{a_{21}}\\
	+{a_{12}}+{a_{22}}\\
	+{a_{33}}-{a_{43}}\\
	-{a_{34}}+{a_{44}}
	\end{array}\right)\tensorproduct
	\left(\begin{array}{c}
	-{b_{11}}-{b_{21}}-{b_{31}}+{b_{41}}\\
	+{b_{12}}+{b_{22}}-{b_{32}}+{b_{42}}\\
	+{b_{13}}+{b_{23}}-{b_{33}}+{b_{43}}\\
	+{b_{14}}+{b_{24}}-{b_{34}}+{b_{44}}
	\end{array}\right)\tensorproduct
	\lfrac{\left(\begin{array}{c}
	-{c_{11}}+{c_{12}}-{c_{13}}+{c_{14}}\\
	-{c_{21}}+{c_{22}}-{c_{23}}+{c_{24}}\\
	-{c_{31}}+{c_{32}}-{c_{33}}-{c_{34}}\\
	+{c_{41}}-{c_{42}}+{c_{43}}+{c_{44}}
	\end{array}\right)}{8},
	\end{aligned}\\
m_{36}=&\begin{aligned}[t]
	\left(\begin{array}{c}
	{a_{11}}+{a_{21}}+{a_{31}}+{a_{41}}
	-{a_{12}}-{a_{22}}-{a_{32}}-{a_{42}}\\
	+{a_{13}}+{a_{23}}+{a_{33}}+{a_{43}}
	+{a_{14}}+{a_{24}}+{a_{34}}+{a_{44}}
	\end{array}\right)\tensorproduct
	\left({b_{12}}+{b_{32}}+{b_{13}}+{b_{33}}\right)\tensorproduct
	\frac{\left({c_{23}}+{c_{43}}\right)}{4},
	\end{aligned}\\
m_{37}=&\begin{aligned}[t]
	\left({a_{12}}+{a_{22}}+{a_{32}}-{a_{42}}\right)\tensorproduct
	\left({b_{22}}+{b_{42}}+{b_{23}}+{b_{43}}+{b_{24}}+{b_{44}}\right)\tensorproduct
	\lfrac{\left({c_{21}}+{c_{23}}-{c_{41}}-{c_{43}}\right)}{2},
	\end{aligned}\\
m_{38}=&\begin{aligned}[t]
	\left(\begin{array}{c}
	{a_{12}}+{a_{22}}\\
	+{a_{32}}-{a_{42}}\\
	+{a_{14}}-{a_{24}}\\
	+{a_{34}}+{a_{44}}
	\end{array}\right)\tensorproduct
	\left(\begin{array}{c}
	-{b_{21}}+{b_{42}}\\
	+{b_{23}}+{b_{43}}\\
	+{b_{24}}+{b_{44}}
	\end{array}\right)\tensorproduct
	\frac{\left(\begin{array}{c}
	{c_{11}}-{c_{12}}+{c_{13}}+{c_{14}}-{c_{21}}-{c_{22}}-{c_{23}}+{c_{24}}\\
	+{c_{31}}-{c_{32}}+{c_{33}}+{c_{34}}+{c_{41}}+{c_{42}}+{c_{43}}-{c_{44}}
	\end{array}\right)}{4},
	\end{aligned}\\
m_{39}=&\begin{aligned}[t]
	\left({a_{12}}+{a_{22}}-{a_{32}}+{a_{42}}\right)\tensorproduct
	\left({b_{41}}-{b_{21}}\right)\tensorproduct
	\lfrac{\left(-{c_{11}}-{c_{13}}+{c_{31}}+{c_{33}}\right)}{2},
	\end{aligned}
	\end{align} \begin{align}
m_{40}=&\begin{aligned}[t]
	\left(\begin{array}{c}
	{-a_{11}}+{a_{21}}+{a_{31}}-{a_{41}}\\
	-{a_{12}}+{a_{22}}-{a_{32}}+{a_{42}}
	\end{array}\right)\tensorproduct
	\left(\begin{array}{c}
		{b_{11}}-{b_{14}}\\
		+{b_{22}}+{b_{24}}
	\end{array}\right)\tensorproduct
	\lfrac{\left(\begin{array}{c}
	2({c_{21}}-{c_{11}}) -{c_{13}}-{c_{14}} +{c_{23}} \\
	+{c_{24}}+{c_{33}}-{c_{34}}-{c_{43}}+{c_{44}}
	\end{array}\right)}{4}
	\end{aligned}\\
m_{41}=&\begin{aligned}[t]
	\left(\begin{array}{c}
	-{a_{11}}-{a_{21}}-{a_{31}}-{a_{41}}\\
	+{a_{12}}+{a_{22}}+{a_{32}}+{a_{42}}\\
	+{a_{13}}+{a_{23}}+{a_{33}}+{a_{43}}\\
	+{a_{14}}+{a_{24}}+{a_{34}}+{a_{44}}
	\end{array}\right)\tensorproduct
	\left(\begin{array}{c}
	{b_{12}}-{b_{32}}\\
	+{b_{13}}-{b_{33}}
	\end{array}\right)\tensorproduct
	\lfrac{\left(\begin{array}{c}
	{c_{11}}-{c_{12}}+{c_{14}}\\
	{c_{31}}-{c_{32}}+{c_{34}}
	\end{array}\right)}{4},
	\end{aligned}\\
m_{42}=&\begin{aligned}[t]
	\left({a_{14}}-{a_{24}}+{a_{34}}-{a_{44}}\right)\tensorproduct
	\left({b_{31}}+{b_{41}}-{b_{34}}-{b_{44}}\right)\tensorproduct
	\lfrac{\left({c_{32}}-{c_{34}}-{c_{42}}+{c_{44}}\right)}{2},
	\end{aligned}\\
m_{43}=&\begin{aligned}[t]
	\left(-{a_{11}}-{a_{21}}+{a_{31}}-{a_{41}}\right)\tensorproduct
	\left({b_{11}}+{b_{31}}\right)\tensorproduct
	\lfrac{\left({c_{21}}+{c_{24}}+{c_{41}}+{c_{44}}\right)}{2},
	\end{aligned}\\
m_{44}=&\begin{aligned}[t]
	\left(\begin{array}{c}
	{a_{11}}+{a_{21}}\\
	+{a_{31}}-{a_{41}}\\
	+{a_{13}}-{a_{23}}\\
	+{a_{33}}+{a_{43}}
	\end{array}\right)\tensorproduct
	\left(\begin{array}{c}
	{b_{11}}+{b_{32}}\\
	-{b_{13}}+{b_{33}}\\
	-{b_{14}}+{b_{34}}
	\end{array}\right)\tensorproduct
	\lfrac{\left(\begin{array}{c}
	{c_{11}}+{c_{12}}-{c_{13}}+{c_{14}}-{c_{21}}\\
	+{c_{22}}-{c_{23}}-{c_{24}}+{c_{31}}+{c_{32}}\\
	-{c_{33}}+{c_{34}}+{c_{41}}-{c_{42}}+{c_{43}}+{c_{44}}
	\end{array}\right)}{4},
	\end{aligned}\\
m_{45}=&\begin{aligned}[t]
	\left(\begin{array}{c}
	{a_{11}}+{a_{21}}-{a_{31}}+{a_{41}}
	+{a_{12}}+{a_{22}}-{a_{32}}+{a_{42}}\\
	-{a_{13}}-{a_{23}}+{a_{33}}-{a_{43}}
	+{a_{14}}+{a_{24}}-{a_{34}}+{a_{44}}
	\end{array}\right)\tensorproduct
	\left( {b_{41}} -{b_{31}} \right)\tensorproduct
	\frac{\left(\begin{array}{c}
		{c_{11}} -{c_{12}} +{c_{13}}\\
		+{c_{21}} -{c_{22}} +{c_{23}}
	\end{array}\right)}{4},
	\end{aligned}\\
m_{46}=&\begin{aligned}[t]
	\left(\begin{array}{c}
	{a_{11}}+{a_{21}}+{a_{31}}-{a_{41}}
	+{a_{12}}+{a_{22}}+{a_{32}}-{a_{42}}\\
	+{a_{13}}+{a_{23}}+{a_{33}}-{a_{43}}
	-{a_{14}}-{a_{24}}-{a_{34}}+{a_{44}}
	\end{array}\right)\tensorproduct
	\left(\begin{array}{c}
	{b_{32}} +{b_{33}} +{b_{34}} \\
	-{b_{42}} -{b_{43}} -{b_{44}}
		\end{array}\right)\tensorproduct
	\frac{\left({c_{14}}+{c_{24}}\right)}{4},
	\end{aligned}\\
m_{47}=&\begin{aligned}[t]
	\left(-{a_{11}}-{a_{21}}+{a_{31}}+{a_{41}}\right)\tensorproduct
	\left({b_{11}}+{b_{21}}-{b_{13}}-{b_{23}}\right)\tensorproduct
	\lfrac{\left({c_{31}}+{c_{34}}+{c_{41}}+{c_{44}}\right)}{2},
	\end{aligned}\\
m_{48}=&\begin{aligned}[t]
	\left(\begin{array}{c}
		-{a_{12}}-{a_{22}}\\
		+{a_{32}}-{a_{42}}\\
		-{a_{14}}+{a_{24}}\\
		+{a_{34}}+{a_{44}}
	\end{array}\right)\tensorproduct
	\left({b_{41}}+{b_{22}}\right)\tensorproduct
	\lfrac{\left(\begin{array}{c}
	-{c_{11}}-{c_{12}}-{c_{13}}+{c_{14}}
	+{c_{21}}-{c_{22}}+{c_{23}}+{c_{24}}\\
	+{c_{31}}+{c_{32}}+{c_{33}}-{c_{34}}
	+{c_{41}}-{c_{42}}+{c_{43}}+{c_{44}}
	\end{array}\right)}{4}.
	\end{aligned}
	\label{eq:lastprod}
\end{align}
To translate this theoretical tensor decomposition into a practical sequence of computational steps, we express it as linear operators a.k.a.~\textsc{lrp} representation (their machine-readable format are given in~\cref{sec:source} and in~\cite{Sedoglavic:FMMDB,jgd:2024:plinopt}).
\section{LRP representation of the algorithm}\label{sec:LRPrepresentation}
While the tensor decomposition representation introduced in the previous section provides a concise mathematical encoding of the algorithm, it does not directly describe the computational steps required for implementation.
To bridge this gap, we now translate the tensor decomposition into the \textsc{lrp} representation.
This framework explicitly decomposes the matrix product into a sequence of linear combinations of input entries, followed by~\(r\) scalar multiplications, and a final set of linear combinations to reconstruct the output matrix.
This representation is the foundation for the numerical schemes and complexity analysis presented in the following sections.
\begin{notation}
Given an~\(\matrixsize{\firstdim}{\seconddim}\)-matrix~\(\mat{A}\), we denote by~\(\row{\mat{A}}{i}\) the~\(i\)-th row and by~\(\vectorization{\mat{A}}\) the row-major vectorization of this matrix, i.e.\ the vector~\(v\) in~\(\K^{\firstdim\seconddim}\) of the entries of the matrix such that~\({v_{i\seconddim+j} = a_{i,j}}\) where indices~\({i,j}\) are~\(0\)-based.
\end{notation}
\begin{definition}\label{def:LRPRepresentation}
The \emph{\textsc{lrp}} representation of a tensor decomposition~\({\sum_{i=1}^{r}{\LeftTensor{i}}\tensorproduct{\RightTensor{i}}\tensorproduct{\ProductTensor{i}}}\) in the tensor product space~\({{\BDual{\Matrices{\K}{\matrixsize{\firstdim}{\seconddim}}}}\tensorproduct{\BDual{\Matrices{\K}{\matrixsize{\seconddim}{\thirddim}}}}\tensorproduct{\Matrices{\K}{\matrixsize{\firstdim}{\thirddim}}}}\), is formed by the three matrices~\(\mat{L}\) in~\(\Matrices{\K}{\matrixsize{r}{\firstdim\seconddim}}\),~\(\mat{R}\) in~\(\Matrices{\K}{\matrixsize{r}{\seconddim\thirddim}}\), and~\(\mat{P}\) in~\(\Matrices{\K}{\matrixsize{\firstdim\thirddim}{r}}\), such that~\(\row{\mat{L}}{i}\) is~\(\vectorization{\LeftTensor{i}}\),~\(\row{\mat{R}}{i}\) is~\(\vectorization{\RightTensor{i}}\) and~\(\col{\mat{P}}{i}\) is~\(\vectorization{\Transpose{\ProductTensor{i}}}\).
\end{definition}
The action of isotropies on an \textsc{lrp} representation is a direct consequence of results presented in~\Cref{sec:Isotropies} and is summarized in the following lemma.
\begin{lemma}\label{lem:actionOnLRPRepresentation}
Let~\(\Isotropy{g}\) be defined by three matrices~\({({\mat{U}},{\mat{V}},{\mat{W}})}\) in~\({{{\emph{\textsc{psl}}^{\pm}({{\K}^{\firstdim}})}}\times{{\emph{\textsc{psl}}^{\pm}({{\K}^{\seconddim}})}}\times{{\emph{\textsc{psl}}^{\pm}({{\K}^{\thirddim}})}}}\) and an \emph{\textsc{lrp}} representation~\(\LRPRepresentation{\mat{L}}{\mat{R}}{\mat{P}}\) of a matrix product tensor decomposition, the action~\({\IsotropyAction{\Isotropy{g}}{\LRPRepresentation{\mat{L}}{\mat{R}}{\mat{P}}}}\) of~\(\Isotropy{g}\) on~\(\LRPRepresentation{\mat{L}}{\mat{R}}{\mat{P}}\) is another \emph{\textsc{lrp}} representation of a matrix product tensor decomposition defined by:
\begin{equation}\label{eq:isotropyActionOnLRPRepresentation}
\LRPRepresentation%
{\MatrixProduct{L}{\bigl({\Transpose{\mat{V}}\tensorproduct{\Inverse{\mat{U}}}}\bigr)}}%
{\MatrixProduct{R}{\bigl({\Transpose{\mat{W}}\tensorproduct{\Inverse{\mat{V}}}}\bigr)}}%
{\MatrixProduct{\bigl({{\mat{U}}\tensorproduct{\InvTranspose{\mat{W}}}}\bigr)}{P}}.
\end{equation}
\end{lemma}
We present now an \textsc{lrp} representation of the tensor decomposition given by~\Crefrange{eq:firstprod}{eq:lastprod} (up to multiplications by~\(2\) of the rows of~\(\mat{L}\) numbered~\({2, 6, 9, 12, 14, 22, 36, 37, 38, 42, 44, 47}\)---in~\(0\)-based indexing, and to divisions by~\(2\) of the columns of~\(\mat{P}\) with the same numbers, thanks to properties of the projective space):
\begin{equation}\label{eq:444LR}
\resizebox{.9\linewidth}{!}{\(%
\mat{L}=\begin{smatrix}
-1&1&1&1&-1&1&1&1&1&-1&-1&-1&-1&1&1&1\\
-1&0&0&0&1&0&0&0&1&0&0&0&-1&0&0&0\\
0&0&-\lfrac{1}{2}&0&0&0&\lfrac{1}{2}&0&0&0&\lfrac{1}{2}&0&0&0&\lfrac{1}{2}&0\\
0&0&-1&-1&0&0&-1&-1&0&0&1&-1&0&0&1&-1\\
1&1&1&-1&-1&-1&-1&1&1&1&1&-1&1&1&1&-1\\
1&1&-1&1&-1&-1&1&-1&-1&-1&1&-1&1&1&-1&1\\
0&0&0&-\lfrac{1}{2}&0&0&0&\lfrac{1}{2}&0&0&0&\lfrac{1}{2}&0&0&0&\lfrac{1}{2}\\
1&1&1&-1&-1&-1&-1&1&1&1&1&-1&-1&-1&-1&1\\
0&0&-1&-1&0&0&1&1&-1&1&0&0&-1&1&0&0\\
\lfrac{1}{2}&0&0&0&\lfrac{1}{2}&0&0&0&\lfrac{1}{2}&0&0&0&-\lfrac{1}{2}&0&0&0\\
-1&-1&-1&1&1&1&1&-1&1&1&1&-1&-1&-1&-1&1\\
0&1&0&1&-1&0&1&0&0&1&0&1&-1&0&1&0\\
0&0&\lfrac{1}{2}&0&0&0&-\lfrac{1}{2}&0&0&0&\lfrac{1}{2}&0&0&0&\lfrac{1}{2}&0\\
0&0&1&0&0&0&1&0&0&0&-1&0&0&0&-1&0\\
\lfrac{1}{2}&0&\lfrac{1}{2}&0&\lfrac{1}{2}&0&-\lfrac{1}{2}&0&-\lfrac{1}{2}&0&-\lfrac{1}{2}&0&\lfrac{1}{2}&0&-\lfrac{1}{2}&0\\
1&1&-1&1&-1&-1&1&-1&1&1&-1&1&-1&-1&1&-1\\
1&1&0&0&1&1&0&0&-1&1&0&0&-1&1&0&0\\
-1&1&1&1&-1&1&1&1&1&-1&-1&-1&1&-1&-1&-1\\
0&1&0&1&-1&0&1&0&0&-1&0&-1&1&0&-1&0\\
-1&1&-1&-1&-1&1&-1&-1&1&-1&1&1&1&-1&1&1\\
0&0&1&0&0&0&-1&0&0&0&-1&0&0&0&1&0\\
1&0&1&0&0&-1&0&1&1&0&1&0&0&-1&0&1\\
0&0&0&\lfrac{1}{2}&0&0&0&-\lfrac{1}{2}&0&0&0&\lfrac{1}{2}&0&0&0&\lfrac{1}{2}\\
1&-1&1&1&-1&1&-1&-1&1&-1&1&1&1&-1&1&1\\
1&1&0&0&-1&-1&0&0&0&0&1&-1&0&0&1&-1\\
1&-1&1&1&1&-1&1&1&1&-1&1&1&-1&1&-1&-1\\
0&1&0&0&0&1&0&0&0&1&0&0&0&1&0&0\\
0&0&1&1&0&0&1&1&1&-1&0&0&-1&1&0&0\\
1&-1&-1&-1&-1&1&1&1&-1&1&1&1&-1&1&1&1\\
0&0&0&1&0&0&0&1&0&0&0&1&0&0&0&1\\
0&1&0&0&0&-1&0&0&0&1&0&0&0&-1&0&0\\
1&0&1&0&0&-1&0&1&-1&0&-1&0&0&1&0&-1\\
1&1&-1&1&-1&-1&1&-1&-1&-1&1&-1&-1&-1&1&-1\\
0&0&-1&-1&0&0&1&1&0&0&1&-1&0&0&-1&1\\
1&1&0&0&1&1&0&0&0&0&1&-1&0&0&-1&1\\
1&-1&1&1&1&-1&1&1&1&-1&1&1&1&-1&1&1\\
0&\lfrac{1}{2}&0&0&0&\lfrac{1}{2}&0&0&0&\lfrac{1}{2}&0&0&0&-\lfrac{1}{2}&0&0\\
0&\lfrac{1}{2}&0&\lfrac{1}{2}&0&\lfrac{1}{2}&0&-\lfrac{1}{2}&0&\lfrac{1}{2}&0&\lfrac{1}{2}&0&-\lfrac{1}{2}&0&\lfrac{1}{2}\\
0&\lfrac{1}{2}&0&0&0&\lfrac{1}{2}&0&0&0&-\lfrac{1}{2}&0&0&0&\lfrac{1}{2}&0&0\\
-1&-1&0&0&1&1&0&0&1&-1&0&0&-1&1&0&0\\
-1&1&1&1&-1&1&1&1&-1&1&1&1&-1&1&1&1\\
0&0&0&1&0&0&0&-1&0&0&0&1&0&0&0&-1\\
\lfrac{1}{2}&0&0&0&\lfrac{1}{2}&0&0&0&-\lfrac{1}{2}&0&0&0&\lfrac{1}{2}&0&0&0\\
\frac{1}{2}&0&\frac{1}{2}&0&\frac{1}{2}&0&-\frac{1}{2}&0&\frac{1}{2}&0&\frac{1}{2}&0&-\frac{1}{2}&0&\frac{1}{2}&0\\
1&1&-1&1&1&1&-1&1&-1&-1&1&-1&1&1&-1&1\\
1&1&1&-1&1&1&1&-1&1&1&1&-1&-1&-1&-1&1\\
-1&0&0&0&-1&0&0&0&1&0&0&0&1&0&0&0\\
0&-\frac{1}{2}&0&-\frac{1}{2}&0&-\frac{1}{2}&0&\frac{1}{2}&0&\frac{1}{2}&0&\frac{1}{2}&0&-\frac{1}{2}&0&\frac{1}{2}\\
 \end{smatrix},\
\mat{R}=\begin{smatrix}
0&0&0&0&0&0&0&0&1&0&0&0&1&0&0&0\\
0&1&0&1&0&-1&0&-1&0&0&0&0&0&0&0&0\\
0&1&0&0&0&0&0&0&0&1&0&0&0&0&0&0\\
0&0&0&0&0&0&0&0&0&1&1&0&1&0&-1&0\\
-1&0&1&1&-1&0&1&1&0&0&0&0&0&0&0&0\\
0&0&0&0&0&1&0&1&0&0&0&0&0&1&0&1\\
0&0&0&0&0&1&0&0&0&0&0&0&0&-1&0&0\\
1&0&0&-1&0&0&0&0&1&0&0&-1&0&0&0&0\\
1&1&-1&-1&-1&-1&1&1&-1&1&1&1&-1&1&1&1\\
0&1&1&1&0&0&0&0&0&-1&-1&-1&0&0&0&0\\
0&0&0&0&0&1&0&1&0&0&0&0&0&-1&0&-1\\
-1&-1&-1&1&1&-1&-1&-1&1&1&1&-1&1&-1&-1&-1\\
1&0&-1&-1&0&0&0&0&-1&0&1&1&0&0&0&0\\
0&0&0&0&0&0&0&0&1&0&-1&0&-1&0&1&0\\
0&1&0&0&0&0&0&0&-1&0&0&0&0&0&0&0\\
1&0&0&-1&0&0&0&0&-1&0&0&1&0&0&0&0\\
0&-1&-1&0&1&0&-1&0&0&0&0&0&0&0&0&0\\
0&0&0&0&1&0&-1&0&0&0&0&0&1&0&-1&0\\
1&-1&-1&-1&1&1&-1&1&-1&1&1&1&1&1&-1&1\\
0&0&0&0&-1&0&1&0&0&0&0&0&1&0&-1&0\\
0&0&0&0&0&0&0&0&0&1&0&1&0&1&0&1\\
1&1&1&-1&-1&1&1&1&1&1&1&-1&1&-1&-1&-1\\
0&0&0&0&-1&0&1&1&0&0&0&0&-1&0&1&1\\
1&0&-1&-1&-1&0&1&1&0&0&0&0&0&0&0&0\\
1&1&-1&-1&1&1&-1&-1&-1&1&1&1&1&-1&-1&-1\\
0&0&0&0&0&0&0&0&0&1&1&1&0&1&1&1\\
0&1&1&0&0&1&1&0&0&0&0&0&0&0&0&0\\
1&-1&-1&-1&-1&1&1&1&1&1&1&1&1&1&1&1\\
0&-1&0&0&0&1&0&0&0&0&0&0&0&0&0&0\\
0&0&0&0&0&0&0&0&0&-1&-1&0&0&1&1&0\\
1&0&0&-1&-1&0&0&1&0&0&0&0&0&0&0&0\\
1&-1&-1&-1&1&1&-1&1&1&-1&-1&-1&-1&-1&1&-1\\
0&1&0&0&0&1&0&0&0&0&0&0&0&0&0&0\\
0&0&0&0&0&0&0&0&-1&0&0&1&0&1&0&1\\
-1&1&1&1&-1&1&1&1&-1&-1&-1&-1&1&1&1&1\\
0&1&1&0&0&0&0&0&0&1&1&0&0&0&0&0\\
0&0&0&0&0&1&1&1&0&0&0&0&0&1&1&1\\
0&0&0&0&-1&0&1&1&0&0&0&0&0&1&1&1\\
0&0&0&0&-1&0&0&0&0&0&0&0&1&0&0&0\\
1&0&0&-1&0&1&0&1&0&0&0&0&0&0&0&0\\
0&1&1&0&0&0&0&0&0&-1&-1&0&0&0&0&0\\
0&0&0&0&0&0&0&0&1&0&0&-1&1&0&0&-1\\
1&0&0&0&0&0&0&0&1&0&0&0&0&0&0&0\\
1&0&-1&-1&0&0&0&0&0&1&1&1&0&0&0&0\\
0&0&0&0&0&0&0&0&-1&0&0&0&1&0&0&0\\
0&0&0&0&0&0&0&0&0&1&1&1&0&-1&-1&-1\\
1&0&-1&0&1&0&-1&0&0&0&0&0&0&0&0&0\\
0&0&0&0&0&1&0&0&0&0&0&0&1&0&0&0\\
\end{smatrix}\)}
\end{equation}
\par
\begin{equation}\label{eq:444P}
\resizebox{.9\linewidth}{!}{\(%
\mat{P}=\begin{smatrix}
0&-\frac{1}{2}&0&0&0&0&0&0&\frac{1}{8}&1&0&\frac{1}{4}&0&0&-\frac{1}{2}&\frac{1}{4}&0&0&0&0&0&0&0&0&-\frac{1}{8}&0&0&\frac{1}{8}&\frac{1}{4}&0&-\frac{1}{2}&0&0&0&-\frac{1}{8}&0&0&\frac{1}{2}&-1&-\frac{1}{2}&\frac{1}{4}&0&0&\frac{1}{2}&\frac{1}{4}&0&0&-\frac{1}{2}\\
0&0&0&-\frac{1}{2}&0&0&1&0&-\frac{1}{8}&0&0&-\frac{1}{4}&-1&\frac{1}{2}&\frac{1}{2}&-\frac{1}{4}&0&0&0&0&0&0&0&0&\frac{1}{8}&0&0&-\frac{1}{8}&-\frac{1}{4}&\frac{1}{2}&0&0&0&0&\frac{1}{8}&0&0&-\frac{1}{2}&0&0&-\frac{1}{4}&0&0&\frac{1}{2}&-\frac{1}{4}&0&0&-\frac{1}{2}\\
0&0&0&\frac{1}{4}&0&0&0&0&-\frac{1}{8}&0&0&\frac{1}{8}&1&-\frac{1}{2}&-\frac{1}{2}&\frac{1}{4}&-\frac{1}{4}&0&-\frac{1}{8}&\frac{1}{4}&0&\frac{1}{8}&0&-\frac{1}{4}&-\frac{1}{8}&0&0&\frac{1}{8}&0&0&-\frac{1}{2}&-\frac{1}{8}&0&\frac{1}{4}&-\frac{1}{8}&0&0&\frac{1}{2}&-1&-\frac{1}{4}&0&0&0&-\frac{1}{2}&\frac{1}{4}&0&0&-\frac{1}{2}\\
0&-\frac{1}{2}&0&\frac{1}{4}&0&0&-1&0&\frac{1}{8}&1&-\frac{1}{4}&\frac{1}{8}&0&0&-\frac{1}{2}&0&\frac{1}{4}&0&\frac{1}{8}&0&0&-\frac{1}{8}&0&0&-\frac{1}{8}&0&0&\frac{1}{8}&\frac{1}{4}&-\frac{1}{2}&0&-\frac{1}{8}&0&-\frac{1}{4}&\frac{1}{8}&0&0&\frac{1}{2}&0&-\frac{1}{4}&\frac{1}{4}&0&0&\frac{1}{2}&0&\frac{1}{4}&0&\frac{1}{2}\\
0&\frac{1}{2}&0&0&0&\frac{1}{4}&0&0&-\frac{1}{8}&0&0&0&0&0&\frac{1}{2}&0&0&\frac{1}{4}&-\frac{1}{4}&0&0&0&0&0&\frac{1}{8}&0&0&\frac{1}{8}&-\frac{1}{4}&0&\frac{1}{2}&0&0&0&-\frac{1}{8}&0&1&-\frac{1}{2}&0&\frac{1}{2}&0&0&1&-\frac{1}{2}&\frac{1}{4}&0&0&\frac{1}{2}\\
0&0&1&-\frac{1}{2}&0&-\frac{1}{4}&0&0&\frac{1}{8}&0&0&0&0&\frac{1}{2}&\frac{1}{2}&0&0&-\frac{1}{4}&\frac{1}{4}&0&0&0&1&0&-\frac{1}{8}&0&0&-\frac{1}{8}&\frac{1}{4}&\frac{1}{2}&0&0&0&0&\frac{1}{8}&0&0&-\frac{1}{2}&0&0&0&0&0&\frac{1}{2}&-\frac{1}{4}&0&0&-\frac{1}{2}\\
0&0&-1&\frac{1}{4}&0&\frac{1}{4}&0&0&\frac{1}{8}&0&0&\frac{1}{8}&0&-\frac{1}{2}&-\frac{1}{2}&0&-\frac{1}{4}&0&-\frac{1}{8}&0&0&-\frac{1}{8}&0&\frac{1}{4}&\frac{1}{8}&0&0&\frac{1}{8}&0&0&\frac{1}{2}&\frac{1}{8}&0&-\frac{1}{4}&-\frac{1}{8}&\frac{1}{4}&1&-\frac{1}{2}&0&\frac{1}{4}&0&0&0&-\frac{1}{2}&\frac{1}{4}&0&0&\frac{1}{2}\\
0&\frac{1}{2}&0&\frac{1}{4}&0&0&0&\frac{1}{4}&-\frac{1}{8}&0&0&-\frac{1}{8}&0&0&\frac{1}{2}&0&\frac{1}{4}&\frac{1}{4}&-\frac{1}{8}&0&0&-\frac{1}{8}&-1&0&\frac{1}{8}&0&0&\frac{1}{8}&-\frac{1}{4}&-\frac{1}{2}&0&-\frac{1}{8}&0&\frac{1}{4}&\frac{1}{8}&0&0&\frac{1}{2}&0&\frac{1}{4}&0&0&1&-\frac{1}{2}&0&\frac{1}{4}&0&\frac{1}{2}\\
-\frac{1}{4}&0&0&0&0&0&0&0&\frac{1}{8}&1&0&\frac{1}{4}&0&0&\frac{1}{2}&\frac{1}{4}&\frac{1}{2}&0&0&0&0&0&0&0&-\frac{1}{8}&0&\frac{1}{2}&\frac{1}{8}&0&0&0&0&\frac{1}{4}&0&-\frac{1}{8}&0&0&\frac{1}{2}&1&0&\frac{1}{4}&0&0&\frac{1}{2}&0&0&\frac{1}{2}&\frac{1}{2}\\
\frac{1}{4}&0&0&0&0&0&-1&0&-\frac{1}{8}&0&0&-\frac{1}{4}&-1&0&-\frac{1}{2}&-\frac{1}{4}&0&0&0&0&-\frac{1}{2}&0&0&0&\frac{1}{8}&0&0&-\frac{1}{8}&0&0&0&0&-\frac{1}{4}&-\frac{1}{2}&\frac{1}{8}&0&0&-\frac{1}{2}&0&0&-\frac{1}{4}&\frac{1}{2}&0&\frac{1}{2}&0&0&0&\frac{1}{2}\\
0&0&0&\frac{1}{4}&0&0&0&0&\frac{1}{8}&0&0&\frac{1}{8}&1&0&\frac{1}{2}&\frac{1}{4}&\frac{1}{4}&0&\frac{1}{8}&-\frac{1}{4}&\frac{1}{2}&\frac{1}{8}&0&0&-\frac{1}{8}&\frac{1}{4}&\frac{1}{2}&-\frac{1}{8}&0&0&0&\frac{1}{8}&\frac{1}{4}&\frac{1}{4}&-\frac{1}{8}&0&0&\frac{1}{2}&1&\frac{1}{4}&0&0&0&-\frac{1}{2}&0&0&0&\frac{1}{2}\\
-\frac{1}{4}&0&0&-\frac{1}{4}&\frac{1}{4}&0&1&0&\frac{1}{8}&1&\frac{1}{4}&\frac{1}{8}&0&0&\frac{1}{2}&0&\frac{1}{4}&0&-\frac{1}{8}&0&0&-\frac{1}{8}&0&0&\frac{1}{8}&0&0&\frac{1}{8}&0&0&0&\frac{1}{8}&0&\frac{1}{4}&-\frac{1}{8}&0&0&\frac{1}{2}&0&-\frac{1}{4}&\frac{1}{4}&-\frac{1}{2}&0&\frac{1}{2}&0&0&\frac{1}{2}&-\frac{1}{2}\\
\frac{1}{4}&0&0&0&0&-\frac{1}{4}&0&0&\frac{1}{8}&0&0&0&0&0&\frac{1}{2}&0&\frac{1}{2}&-\frac{1}{4}&\frac{1}{4}&0&0&0&0&0&-\frac{1}{8}&0&\frac{1}{2}&-\frac{1}{8}&0&0&0&0&\frac{1}{4}&0&\frac{1}{8}&0&-1&\frac{1}{2}&0&0&0&0&1&\frac{1}{2}&0&0&\frac{1}{2}&\frac{1}{2}\\
-\frac{1}{4}&0&1&0&0&\frac{1}{4}&0&0&-\frac{1}{8}&0&0&0&0&0&\frac{1}{2}&0&0&\frac{1}{4}&-\frac{1}{4}&0&\frac{1}{2}&0&-1&0&\frac{1}{8}&0&0&\frac{1}{8}&0&0&0&0&-\frac{1}{4}&\frac{1}{2}&-\frac{1}{8}&0&0&\frac{1}{2}&0&0&0&-\frac{1}{2}&0&-\frac{1}{2}&0&0&0&-\frac{1}{2}\\
0&0&-1&\frac{1}{4}&0&-\frac{1}{4}&0&0&\frac{1}{8}&0&0&\frac{1}{8}&0&0&-\frac{1}{2}&0&\frac{1}{4}&0&\frac{1}{8}&0&-\frac{1}{2}&-\frac{1}{8}&0&0&-\frac{1}{8}&-\frac{1}{4}&\frac{1}{2}&\frac{1}{8}&0&0&0&-\frac{1}{8}&\frac{1}{4}&-\frac{1}{4}&\frac{1}{8}&\frac{1}{4}&-1&\frac{1}{2}&0&-\frac{1}{4}&0&0&0&\frac{1}{2}&0&0&0&\frac{1}{2}\\
\frac{1}{4}&0&0&-\frac{1}{4}&\frac{1}{4}&0&0&\frac{1}{4}&\frac{1}{8}&0&0&-\frac{1}{8}&0&0&\frac{1}{2}&0&\frac{1}{4}&-\frac{1}{4}&\frac{1}{8}&0&0&-\frac{1}{8}&1&0&\frac{1}{8}&0&0&-\frac{1}{8}&0&0&0&\frac{1}{8}&0&-\frac{1}{4}&\frac{1}{8}&0&0&-\frac{1}{2}&0&\frac{1}{4}&0&\frac{1}{2}&1&\frac{1}{2}&0&0&\frac{1}{2}&\frac{1}{2}\\
\end{smatrix}
\)}
\end{equation}
This matrices have respectively $400$, $240$ and $320$ non-zero
coefficients
(including $64$, $0$ and $304$ non-zero coefficients that are neither
$1$, nor $-1$).
A naive derivation of a program from those
would therefore require $3$ matrix-vector multiplications with a total
of $960-48-48-16+368=1216$ operations.
We show thereafter
a program that uses far fewer operations than that.

\section[Straight-line programs for the rational 4x4x4:48 algorithm]{Straight-line programs for the rational~\(\FMMA{4}{4}{4}{48}\) algorithm}\label{sec:NumericalSchemeAndComplexity}
The \plinopt{} software~\cite{jgd:2024:plinopt,Dumas:2026:autoaccurate} produces straight-line programs from an \textsc{lrp} representation, optimizing its arithmetic cost by several techniques of common sub-expression elimination.
The straight-line programs obtained from the~\({\mat{L},\mat{R},\mat{P}}\) matrices presented in~\Cref{sec:LRPrepresentation} and associated to the tensor decomposition given by~\Crefrange{eq:firstprod}{eq:lastprod} are the following:
\par
	\begin{lstlisting}[style=slp,caption={\SLP for input linear forms defined by matrix~\(\mat{L}\) of~\cref{eq:444LR}},label=lst:444L]
x10:=A13+A24; x11:=A14+A23; x12:=A12-A21; x13:=A31-A42; x14:=A33+A44; x15:=A34+A43; x16:=A22-A11; x17:=A32-A41; x18:=A13-A23; x19:=A32-A42; x20:=A33+A43; x21:=A31-A41; x22:=A34+A44; x23:=A12+A22; x24:=A11+A21; x25:=A14-A24; x26:=x17-x13; x27:=x10+x11; x28:=x14-x15; x29:=x16-x12; x30:=x14+x15; x31:=x12+x16; x32:=x10-x11; x33:=x13+x17; x34:=x23+x24; x35:=x19-x21; x36:=x20-x22; x37:=x18+x25; l8:=x26-x37; x38:=A33-A43; x39:=A31+A41; x40:=A13+A23; l34:=x28+x34; l27:=x27-x35; x41:=A32+A42; x42:=A12-A22; x43:=A14+A24; l24:=x36-x29; x44:=A34-A44; x45:=A11-A21; x46:=x11+x12; x47:=x28-x29; x48:=x31+x26; x49:=x32+x39+x41; l36:=(x19+x23)/2; l38:=l36-x19; x50:=x30+x27; l12:=(x18+x20)/2; l2:=l12-x18; x51:=x30-x27; l22:=(x22+x25)/2; l6:=l22-x25; x52:=x28+x29; x53:=x33-x32; x54:=x10-x16; x55:=x43-x33-x40; x56:=x32+x33; x57:=x31-x26; x58:=x45+x30-x42; l9:=(x21+x24)/2; l42:=l9-x21; x59:=x38+x44-x31; x60:=x13+x14; x61:=x15+x17; l0:=l27-x59; l1:=x21-x45; l3:=x36-x27; l4:=l24+x49; l5:=x47-x56; l7:=x47+x56; l10:=x52+x53; l11:=x46+x61; l13:=x40-x20; l14:=l42-l2; l15:=x53-x52; l16:=x34+x26; l17:=x57-x51; l18:=x46-x61; l19:=x51+x57; l20:=x18-x38; l21:=x54+x60; l23:=x58-l8; l25:=l27+x59; l26:=x23+x41; l28:=x58+l8; l29:=x22+x43; l30:=x19+x42; l31:=x54-x60; l32:=l24-x49; l33:=x28-x37; l35:=x50-x48; l37:=l36+l22; l39:=x29-x35; l40:=x48+x50; l41:=x25+x44; l43:=l12+l9; l44:=l34+x55; l45:=l34-x55; l46:=x39-x24; l47:=l6-l38;
\end{lstlisting}
\begin{lstlisting}[style=slp,caption={\SLP for input linear forms defined by matrix~\(\mat{R}\) of~\cref{eq:444LR}},label=lst:444R]
y21:=B21-B23; y24:=B32+B33; y25:=B42+B44; y19:=B11-B14; r38:=B41-B21; y22:=B41-B43; r0:=B41+B31; r44:=B41-B31; r42:=B11+B31; y20:=B31-B34; r6:=B22-B42; r47:=B41+B22; y26:=B22+B24; r32:=B22+B12; r28:=B22-B12; r2:=B32+B12; r14:=B12-B31; y23:=B12+B13; y27:=B13-y19; r15:=y19-y20; r7:=y19+y20; y18:=B24-y21; r17:=y21+y22; r19:=y22-y21; r16:=y21-y23; y30:=B21+B22; r26:=y30-r16; r40:=y23-y24; r3:=y22+y24; y28:=B34+y24; y31:=B42+B43; r29:=y31-y24; r35:=y23+y24; y29:=y31+B44; r33:=y25-y20; r39:=y19+y26; r30:=r39-y30; r5:=y25+y26; y11:=B12+B11; r1:=y11-r39; r46:=y11+r16; r10:=y26-y25; r43:=y28-y27; y17:=r39-r33; y16:=r43-B31; r12:=y16-B32; y14:=r17-r40; y13:=r15+r5; y12:=r3-r16; r11:=y14-y13; r18:=y14+y13; r45:=y28-y29; r25:=y28+y29; r31:=y17-y12; r21:=y17+y12; r37:=y29+y18; y15:=r37-B41; r22:=y15-B42; r36:=y30+r37; r9:=y11-r43; r41:=r0-B44-B34; r23:=y18-y27; r4:=y18+y27; r8:=y16+y15-r28; y10:=B32+B31; r20:=y10+r33; r13:=y10-r3; y32:=r43-r37; y33:=r32+r44; r24:=y32+y33; r34:=y33-y32; r27:=(r36+r38)*2-r34;
\end{lstlisting}
\par\medskip
\begin{lstlisting}[style=slp, caption={Products},label=lst:444H]
p0:=l0*r0; p1:=l1*r1; p2:=l2*r2; p3:=l3*r3; p4:=l4*r4; p5:=l5*r5; p6:=l6*r6; p7:=l7*r7; p8:=l8*r8; p9:=l9*r9; p10:=l10*r10; p11:=l11*r11; p12:=l12*r12; p13:=l13*r13; p14:=l14*r14; p15:=l15*r15; p16:=l16*r16; p17:=l17*r17; p18:=l18*r18; p19:=l19*r19; p20:=l20*r20; p21:=l21*r21; p22:=l22*r22; p23:=l23*r23; p24:=l24*r24; p25:=l25*r25; p26:=l26*r26; p27:=l27*r27; p28:=l28*r28; p29:=l29*r29; p30:=l30*r30; p31:=l31*r31; p32:=l32*r32; p33:=l33*r33; p34:=l34*r34; p35:=l35*r35; p36:=l36*r36; p37:=l37*r37; p38:=l38*r38; p39:=l39*r39; p40:=l40*r40; p41:=l41*r41; p42:=l42*r42; p43:=l43*r43; p44:=l44*r44; p45:=l45*r45; p46:=l46*r46; p47:=l47*r47;
\end{lstlisting}
\par
\begin{lstlisting}[style=slp,caption=\SLP for input linear forms
  defined by matrix~\(\mat{P}\) of~\cref{eq:444P},label=lst:444P]
z33:=(p0+p44)/4; z34:=p24+p4-p32; z32:=(p32-p28)/4; z40:=p47+z33+z32; z31:=(p14-z40)/2; z11:=p13-p46; z44:=p3-p13-p29; z20:=p29+p26; z51:=p16+p46+p26; z30:=(p25+p45)/4; z29:=(p33-p41-z44+p20)/4; z16:=z31+z29; z28:=(p39+p30+p1+z51)/4; z35:=z40+z31+z28; z36:=p8+p28+p23; z37:=p27-p0-p25-z36; z14:=p31-z37; z15:=p42+z35; z58:=z34+p45+p34-p44; z59:=z20+(p11+p21-z58)/2; z60:=p30-p41; z21:=z60-z59; z46:=z15+(p7-z59-z60)/4; z27:=(z58-z37)/2; z17:=z34-z27; z38:=z27-z21-p11-z36; z12:=p2+z16; z39:=p6+z32+z16; z26:=(z38-p40)/4; z41:=p20-p1+(p18+z14)/2; z10:=z11+z41; z25:=(z41-z11-p17)/4; z42:=p38-z33+z35; z43:=z20/2+(p35-z38)/4-z12; z45:=(p10+z14-z10+z17)/4+z39; z47:=(p19-z10)/4-z42; z24:=(p4-p23)/4; z48:=p43-z30+z24+p37; z49:=z29+p37+(z44-z48)/2; z50:=(z51+z48)/2-z28; z23:=(p5-p18+p17+z17)/4; z13:=p22-z49; z52:=p36+z23-z50; z53:=p9+z30-z26+z50; z19:=z25+z13; z54:=z13-z23; z18:=z25+z52; C41:=z15-z52; C21:=z15+z52; C42:=z12-z54; C22:=z12+z54; z22:=(p15-z21)/4; z55:=p12+z24+z49+z22; z56:=z26-z55; C24:=z46-z19; C44:=z46+z19; C32:=z56-z39; C12:=z39+z56; C33:=z55-z47; C13:=z55+z47; C43:=z43-z18; C23:=z43+z18; z57:=z22+z53; C31:=z42+z57; C11:=z57-z42; C14:=z53-z45; C34:=z53+z45;
\end{lstlisting}
These straight-line programs require:
\begin{itemize}
\item \(104\) additions and~\(4\) multiplications (binary shifts) for \texttt{L};
\item \(75\) additions and~\(1\) multiplication (binary shift) for \texttt{R};
\item \(110\) additions and~\(21\) multiplications (binary shifts) for \texttt{P}.
\end{itemize}
This gives a total of~\(315\) operations and a theoretical complexity bound of:
\begin{equation}
\left(1+\frac{315}{48-16}\right)n^{2+\log_4{\!3}}-\left(\frac{315}{48-16}\right)n^{2}
\approx{10.84375n^{2.7925} - 9.84375n^{2}}.
\end{equation}
Although~\Crefrange{eq:firstprod}{eq:lastprod} present only~\(1\) and~\(-1\) coefficients for~\(b_{ij}\) (see also matrix~\(\mat{R}\) in~\cref{sec:LRPrepresentation}), the \emph{best} \SLP we found does require some multiplications by non-unit coefficients.
This is due to the fact that we use the kernel decomposition of a linear operator (see~\cite[Alg.~2]{Dumas:2026:autoaccurate}): more precisely this means that for some rows of~\(\mat{R}\), it is more efficient to compute them from linear combinations of other rows, even using non-unit linear combination coefficients.
\par
The program for matrix multiplications
combining~\cref{lst:444L,lst:444R,lst:444H,lst:444P} together with a Maple command for its
verification is embedded within this~\textsc{pdf} article as the text file
\verb!4x4x4_48_check.mpl!\embedfile{4x4x4_48_check.mpl}.
\section{Alternative bases algorithm}\label{sec:AlternativeBasis}
Following~\cite{Karstadt:2017aa,Beniamini:2019aa}, we present in this
section an alternative basis derived from
the tensor decomposition given
by~\Crefrange{eq:firstprod}{eq:lastprod}.
We choose a factorization of each one of the~\({48{\times}16}\)~\({\mat{L},\mat{R},\Transpose{\mat{P}}}\) matrices into~\({48{\times}47}\) by~\({47{\times}16}\) matrix products~\({\mat{X}=\mat{X}_{\textup{alt}} \cdot \mat{X}_{\textup{cob}}}\), for~\(\mat{X}\) in~\({\{\mat{L},\mat{R},\Transpose{\mat{P}}\}}\).
Looking at the \SLP of \cref{lst:444L}, we see that
\texttt{l47:=l6-l38} (and that \texttt{l47} is not reused in this \SLP).
Therefore, in~\(0\)-based numbering, this means that row~\(47\) is the
difference of rows~\(6\) and~\(38\).
Let~\(\mat{L}_{h}\) be the first~\(47\) rows of~\(\mat{L}\)
and~\(l_{47}\) the~\(48\)-th of~\cref{eq:444LR}, left.
Let also~\(\vec{e_i}\) (again~\(0\)-indexed) be the~\(i\)-th vector of the canonical basis.
We obtain the decomposition:
\begin{equation}
{\mat{L}=\begin{smatrix} \mat{L}_h \\ l_{47}\end{smatrix} =
  \mat{L}_{\textup{alt}} \cdot \mat{L}_{\textup{cob}} =
  \begin{smatrix}\mat{I}_{47}\\\vec{e_6}-\vec{e_{38}}\end{smatrix}\cdot\mat{L}_h}.
\end{equation}
Similarly, we see in~\cref{lst:444R} that, e.g., \texttt{r20:=y10+r33}
and \texttt{r13:=y10-r3}, so that \texttt{r20:=r33+r3+r13}. We can thus define accordingly, up to row permutations,~\({\mat{R}_{\textup{alt}}=\begin{smatrix} \mat{I}_{47} \\ \vec{e_3}+\vec{e_{13}}+\vec{e_{33}}\end{smatrix}}\) and~\({\mat{R}_{\textup{cob}}}\) as~\(\mat{R}\) of~\cref{eq:444LR}, right, with its~\(20\)-th row removed.
Finally, looking at~\cref{eq:444P}, there is a linear combination of~\(3\)
columns of~\(\mat{P}\) that gives a fourth
one:~\({c_{47}=c_{43}+c_{39}-c_{15}}\).
We can therefore
define~\({\mat{P}_{\textup{alt}}=\Transpose{\begin{smatrix}
    \mat{I}_{47} \\
    \vec{e_{43}}+\vec{e_{39}}-\vec{e_{15}}\end{smatrix}}}\)
and~\(\mat{P}_{\textup{cob}}\) as~\(\mat{P}\)
of~\cref{eq:444P}, with its last column removed.
\par
The algorithm defined by these~\({\mat{L}_{\textup{alt}},\mat{R}_{\textup{alt}},\mat{P}_{\textup{alt}}}\) matrices, can be realized with straight-line programs with, respectively,~\({1,2}\) and~\(3\) additions (for the latter,~\(\Transpose{\mat{P}}_{\textup{alt}}\) can obviously be realized with~\(2\) additions, and therefore~\(\mat{P}_{\textup{alt}}\) requires~\({2+(48-47)=3}\) additions, by the transposition principle).
This gives a theoretical complexity bound of:
\begin{equation}\label{eq:altbasecomp}
\left(1+\frac{1+2+3}{48-47}\right)n^{\log_4{\!48}}
-\left(\frac{1+2+3}{48-47}\right)n^{\log_4{\!47}}
\approx 7n^{2.7925}.
\end{equation}
The cost induced by the respective change-of-basis matrices~\({\mat{L}_{\textup{cob}}, \mat{R}_{\textup{cob}}}\) and~\(\mat{P}_{\textup{cob}}\) is~\({o\!\left(n^{\log_4{\!48}}\right)}\) for~\cref{eq:altbasecomp}~\cite{Beniamini:2019aa}.
The associated straight-line programs require indeed:
\begin{itemize}
\item \(103\) additions  and~\(4\) multiplications (binary shifts) for~\(\mat{L}_{\textup{cob}}\);
\item \(74\) additions and~\(1\) multiplication (binary shift) for~\(\mat{R}_{\textup{cob}}\);
\item \(109\) additions and~\(21\) multiplications (binary shifts) for~\(\mat{P}_{\textup{cob}}\).
\end{itemize}
This is a total of~\(312\) operations for an added complexity bound
of~\({\left(\frac{312}{47-16}\right)\left(n^{\log_4{\!47}}-n^2\right)=o\!\left(n^{2+\log_4{\!3}}\right)}\).
\section{Conclusion}\label{sec:conclusion}
Recent independent work~\cite{Moran:2026aa} corroborates the results
presented in this paper and extends them to additional examples and
different fields.
For instance, it asserts that the tensor decomposition announced
in~\cite{Kaporin:2024ab} does not belong to the same orbit as the
decomposition studied there and cannot be expressed without complex
numbers.
\par
Transforming the complex-valued fast matrix multiplication algorithms
introduced in~\cite{alphaevolve} into a rational-valued algorithm is a
crucial first step toward demonstrating their practical viability.
Further work is required in several complementary directions: first,
reducing further the length of the straight-line programs will
decrease the leading constant in the asymptotic time complexity of the
resulting recursive matrix multiplication algorithm.
While our proposed \SLP already significantly reduces this constant
compared to a naive derivation from the \textsc{lrp} representation,
preliminary results,
on other orbit points and with additional exploration techniques,
suggest that a refined complexity
of~\({{\frac{262}{32}n^{2+\log_{4}{\!3}}}\approx{8.1875n^{2.7925}}}\)
may be achievable.
Also the numerical accuracy of these variants is studied
in~\cite{Dumas:2026ac} and a theoretically optimal one in terms of
numerical accuracy was obtained following the same approach as
in~\cite{jgd:2024:accurate,Dumas:2026:autoaccurate}.
Lastly, translating a straight-line program into an efficient
practical implementation will require finding an operation schedule
that minimizes the temporary memory allocations and maximizes cache
efficiency.
\bibliographystyle{plainurl}
\bibliography{mm444}
\appendix
\crefalias{section}{appendix} %
\section{Source programs availability}\label{sec:source}
All the different matrices and straight-line programs presented in
this note can be found in the \plinopt~library's data
directory~\cite{jgd:2024:plinopt}:
\begin{itemize}
\item
For the three matrices and three straight-line programs presented in
\Cref{sec:LRPrepresentation}:
\plinoptdata{data/4x4x4_48_rational_{L,R,P}.{sms,slp}};
\item For alternative basis based matrices and straight-line programs
  presented in~\Cref{sec:AlternativeBasis}:
\plinoptdata{data/4x4x4_48_rational-{ALT,CoB}_{L,R,P}.{sms,slp}};
\item For matrices and straight-line programs presented
  in~\Cref{sec:LRPrepresentation63}:
\plinoptdata{data/3x4x7_63_rational_{L,R,P}.{sms,slp}};
\item For alternative basis based matrices and straight-line programs
  presented in~\Cref{sec:AlternativeBasis63}:
\plinoptdata{data/3x4x7_63_rational-{ALT,CoB}_{L,R,P}.{sms,slp}}.
\end{itemize}
\section[A rational 3x4x7:63 algorithm]{A rational~\FMMA{3}{4}{7}{63} algorithm}\label{sec:347}
The success of applying isotropy actions to project complex-valued algorithms onto the rational field is not limited to square~\(\matrixsize{4}{4}\) matrices.
To demonstrate the broader applicability of this methodology, we now examine the rectangular case of multiplying a~\(\matrixsize{3}{4}\) matrix by a~\(\matrixsize{4}{7}\) matrix.
By identifying the relevant symmetries and stabilizers for the AlphaEvolve's~\(63\)-multiplication complex algorithm recently discovered via evolutionary agents, we provide a new, equivalent rational-valued decomposition that maintains the same optimal multiplication count while simplifying the underlying arithmetic requirements.
As was done for the~\(\FMMA{4}{4}{4}{48}\) algorithm, we answer positively to the following question:
\begin{problem}
Is there a tensor decomposition in the orbit of~\(\FMMA{3}{4}{7}{63}\) introduced in~\cite{alphaevolve} that does not require complex numbers?
\end{problem}
\subsection{A rational point in the orbit of AlphaEvolve's \texorpdfstring{\(\FMMA{3}{4}{7}{63}\)}{<3x4x7:63>} tensor decomposition}\label{sec:The-3x4x7:63-Algorithm}
Using the Klein four stabilizer group (see~\cite{Sedoglavic:FMMDB}) of the original algorithm presented in~\cite{alphaevolve} alongside guided algebraic searches similar to those done in~\Cref{sec:Stabilizer}, we identified the following isotropy:
\begin{equation}
	{\begin{bmatrix}
		1&0&0\\
		0&I&0\\
		0&0&1
	\end{bmatrix}}
	\times
	{\begin{bmatrix}
		I&0&0&0\\
		0&I&0&0\\
		0&0&1&0\\
		0&0&0&1
	\end{bmatrix}}
	\times
	{\begin{bmatrix}
		I&0&0&0&0&0&0\\
		0&I&0&0&0&0&0\\
		0&0&1&0&0&0&0\\
		0&0&0&I&0&0&0\\
		0&0&0&0&1&0&0\\
		0&0&0&0&0&I&0\\
		0&0&0&0&0&0&1
	\end{bmatrix}},
\end{equation}
that sends the original complex-valued tensor decomposition to the following rational one, as shown in~\Crefrange{eq:firstprod63}{eq:lastprod63}.
\begin{align}
	\label{eq:firstprod63}
m_{1}=&\begin{aligned}[t]
 \left({a_{21}} +{a_{24}} \right)\tensorproduct\left({b_{17}} -{b_{37}} +{b_{47}} \right)\tensorproduct\left(\frac{{c_{21}}}{2}-\frac{{c_{22}}}{2}+\frac{{c_{24}}}{2}+\frac{{c_{26}}}{2}+\frac{{c_{27}}}{2}\right),
 \end{aligned}\\
m_{2}=&\begin{aligned}[t]
 \left({a_{11}} -{a_{21}} +\frac{{a_{12}}}{2}-\frac{{a_{22}}}{2}\right)\tensorproduct\left( \frac{{b_{32}}}{2}+\frac{{b_{42}}}{2} -\frac{{b_{12}}}{2}-{b_{22}} -{b_{14}} +{b_{15}} +{b_{17}} \right)\tensorproduct\left( {c_{16}} +{c_{17}} -{c_{12}} \right),
 \end{aligned}\\
m_{3}=&\begin{aligned}[t]
 \left(\frac{{a_{32}}}{2} -\frac{{a_{22}}}{2} +{a_{24}} -{a_{34}} \right)\tensorproduct\left(\frac{{b_{11}}}{2}+{b_{21}} +\frac{{b_{31}}}{2}-\frac{{b_{41}}}{2}+{b_{43}} +{b_{46}} -{b_{47}} \right)\tensorproduct\left({c_{31}} +{c_{34}} +{c_{37}} \right),
 \end{aligned}\\
m_{4}=&\begin{aligned}[t]
 \left({a_{11}} +{a_{13}} \right)\tensorproduct\left({b_{32}} +{b_{15}} +\frac{{b_{17}}}{2}+\frac{{b_{37}}}{2}-\frac{{b_{47}}}{2}\right)\tensorproduct\left({c_{12}}+{c_{15}} -{c_{17}} \right),
 \end{aligned}\\
m_{5}=&\begin{aligned}[t]
	\left(\!\begin{array}{c}
		{a_{14}} +{a_{34}}\\
		-\frac{{a_{12}}}{2}-\frac{{a_{32}}}{2}
		\end{array}\!\right)\!\tensorproduct\!\left(\frac{{b_{11}}}{2}+{b_{21}} +\frac{{b_{31}}}{2}-\frac{{b_{41}}}{2}+{b_{43}} +\frac{{b_{14}}}{2}-{b_{24}} +\frac{{b_{34}}}{2}+\frac{{b_{44}}}{2}+{b_{46}} -{b_{47}} \right)\tensorproduct{c_{34}},
 \end{aligned}\\
m_{6}=&\begin{aligned}[t]
 \left({a_{31}} +{a_{34}} \right)\tensorproduct\left({b_{11}} +{b_{46}} -\frac{{b_{17}}}{2}+\frac{{b_{37}}}{2}-\frac{{b_{47}}}{2}\right)\tensorproduct\left({c_{31}} +{c_{36}} +{c_{37}} \right),
 \end{aligned}\\
m_{7}=&\begin{aligned}[t]
 \left({a_{31}} +\frac{{a_{32}}}{2}\right)\tensorproduct\left({b_{16}} -{b_{36}} -{b_{46}} \right)\tensorproduct\left(-{c_{16}} +{c_{26}} +{c_{36}} \right),
 \end{aligned}\\
m_{8}=&\begin{aligned}[t]
 \left({a_{21}} -{a_{31}} +\frac{{a_{22}}}{2}-\frac{{a_{32}}}{2}\right)\tensorproduct\left(\frac{{b_{16}}}{2}+{b_{26}} +\frac{{b_{36}}}{2}+\frac{{b_{46}}}{2}-{b_{17}} \right)\tensorproduct{c_{26}},
 \end{aligned}\\
m_{9}=&\begin{aligned}[t]
 \left(\frac{{a_{22}}}{2}+\frac{{a_{32}}}{2}-{a_{24}}+{a_{34}} \right)\tensorproduct\left(\frac{{b_{11}}}{2}+{b_{21}}+\frac{{b_{31}}}{2}-\frac{{b_{41}}}{2}\right)\tensorproduct\left({c_{21}} +{c_{31}} +{c_{34}} +{c_{37}} \right),
 \end{aligned}\\
m_{10}=&\begin{aligned}[t]
 \left(\frac{{a_{12}}}{2}+\frac{{a_{32}}}{2}-{a_{13}} +{a_{34}} \right)\tensorproduct\left(-{b_{31}} +{b_{45}} -\frac{{b_{17}}}{2}+\frac{{b_{37}}}{2}+\frac{{b_{47}}}{2}\right)\tensorproduct\left({c_{11}} +{c_{35}} \right),
 \end{aligned}\\
m_{11}=&\begin{aligned}[t]
 \left(\frac{{a_{32}}}{2}+{a_{33}} \right)\tensorproduct\left({b_{13}} -{b_{33}} +{b_{43}} \right)\tensorproduct\left({c_{13}} -{c_{23}} -{c_{33}} \right),
 \end{aligned}\\
m_{12}=&\begin{aligned}[t]
 \left(\frac{{a_{32}}}{2} -\frac{{a_{22}}}{2} +{a_{23}} -{a_{33}} \right)\tensorproduct\left({b_{31}} +\frac{{b_{13}}}{2}+{b_{23}} -\frac{{b_{33}}}{2}+\frac{{b_{43}}}{2}+{b_{36}} -{b_{37}} \right)\tensorproduct\left({c_{33}} -{c_{35}} +{c_{37}} \right),
 \end{aligned}\\
m_{13}=&\begin{aligned}[t]
 \left(\frac{{a_{32}}}{2}+{a_{34}} \right)\tensorproduct\left({b_{11}} +{b_{31}} -{b_{41}} \right)\tensorproduct\left({c_{11}} -{c_{21}} -{c_{31}} \right),
 \end{aligned}
\end{align}
\begin{align}
m_{14}=&\begin{aligned}[t]
 \left({a_{21}} -\frac{{a_{12}}}{2}+\frac{{a_{22}}}{2}+{a_{13}} \right)\tensorproduct\left({b_{15}} +\frac{{b_{17}}}{2}+\frac{{b_{37}}}{2}-\frac{{b_{47}}}{2}\right)\tensorproduct\left(-{c_{12}} +{c_{16}} +{c_{17}} +{c_{25}} \right),
 \end{aligned}\\
m_{15}=&\begin{aligned}[t]
 \left({a_{11}} +{a_{14}} \right)\tensorproduct\left({b_{42}} -{b_{14}} +\frac{{b_{17}}}{2}-\frac{{b_{37}}}{2}+\frac{{b_{47}}}{2}\right)\tensorproduct\left({c_{12}} -{c_{14}} -{c_{17}} \right),
 \end{aligned}\\
m_{16}=&\begin{aligned}[t]
 \left(-{a_{21}} +{a_{31}} +\frac{{a_{22}}}{2}+\frac{{a_{32}}}{2}\right)\tensorproduct\left(-\frac{{b_{16}}}{2}+{b_{26}} +\frac{{b_{36}}}{2}+\frac{{b_{46}}}{2}\right)\tensorproduct\left({c_{26}} -{c_{32}} +{c_{36}} +{c_{37}} \right),
 \end{aligned}\\
m_{17}=&\begin{aligned}[t]
\left(\begin{array}{c}{a_{11}} +{a_{31}}\\-\frac{{a_{12}}}{2}+\frac{{a_{32}}}{2}\end{array}\right)\tensorproduct\left(\frac{{b_{12}}}{2}+{b_{22}} -\frac{{b_{32}}}{2}-\frac{{b_{42}}}{2}-\frac{{b_{16}}}{2}+{b_{26}} +\frac{{b_{36}}}{2}+\frac{{b_{46}}}{2}\right)\tensorproduct\left({c_{32}} -{c_{16}} \right),
 \end{aligned}\\
m_{18}=&\begin{aligned}[t]
\left(\!\begin{array}{c}\frac{{a_{12}}}{2}+\frac{{a_{32}}}{2}\\-{a_{13}} -{a_{33}}\end{array}\!\right)\!\tensorproduct\!\left(
	\begin{array}{c}
	{b_{31}} +{b_{36}} -{b_{37}} +{b_{23}} +{b_{25}} \\
		+\lfrac{({b_{13}}-{b_{33}}+{b_{43}}-{b_{15}} -{b_{35}}-{b_{45}})}{2}
	\end{array}
	\right)\tensorproduct{c_{35}},
 \end{aligned}\\
m_{19}=&\begin{aligned}[t]
 \left({a_{11}} -\frac{{a_{12}}}{2}+\frac{{a_{22}}}{2}+{a_{23}} \right)\tensorproduct\left({b_{32}} +\frac{{b_{17}}}{2}+\frac{{b_{37}}}{2}-\frac{{b_{47}}}{2}\right)\tensorproduct\left({c_{13}} -{c_{15}} +{c_{17}} +{c_{22}} \right),
 \end{aligned}\\
m_{20}=&\begin{aligned}[t]
 \left({a_{21}} -{a_{31}} -\frac{{a_{22}}}{2}+\frac{{a_{32}}}{2}\right)\tensorproduct\left({b_{11}} +{b_{13}} -\frac{{b_{16}}}{2}+{b_{26}} +\frac{{b_{36}}}{2}+\frac{{b_{46}}}{2}-{b_{17}} \right)\tensorproduct\left( {c_{36}} -{c_{32}} +{c_{37}} \right),
 \end{aligned}\\
m_{21}=&\begin{aligned}[t]
 \left({a_{13}} +{a_{14}} \right)\tensorproduct\left(-{b_{34}}
 +{b_{45}} -\frac{{b_{17}}}{2} +\frac{{b_{37}}}{2} +\frac{{b_{47}}}{2}\right)\tensorproduct\left(-{c_{14}} +{c_{15}} -{c_{17}} \right),
 \end{aligned}\\
m_{22}=&\begin{aligned}[t]
 \left(\frac{{a_{12}}}{2}+\frac{{a_{32}}}{2}+{a_{33}} -{a_{14}} \right)\tensorproduct\left({b_{43}} +{b_{34}} +\frac{{b_{17}}}{2}-\frac{{b_{37}}}{2}-\frac{{b_{47}}}{2}\right)\tensorproduct\left(-{c_{13}} +{c_{34}} \right),
 \end{aligned}\\
m_{23}=&\begin{aligned}[t]
 \left(-{a_{11}} +{a_{21}} +\frac{{a_{12}}}{2}-\frac{{a_{22}}}{2}\right)\tensorproduct\left(\frac{{b_{12}}}{2}-{b_{22}} +\frac{{b_{32}}}{2}+\frac{{b_{42}}}{2}+{b_{17}} \right)\tensorproduct{c_{22}},
 \end{aligned}\\
m_{24}=&\begin{aligned}[t]
 \left(-{a_{11}} +\frac{{a_{12}}}{2}+\frac{{a_{32}}}{2}+{a_{34}} \right)\tensorproduct\left(-{b_{11}} +{b_{42}} +\frac{{b_{17}}}{2}-\frac{{b_{37}}}{2}+\frac{{b_{47}}}{2}\right)\tensorproduct\left({c_{11}} +{c_{32}} \right),
 \end{aligned}\\
m_{25}=&\begin{aligned}[t]
 \left({a_{21}} -\frac{{a_{22}}}{2} +\frac{{a_{32}}}{2} +{a_{33}}
 \right)\tensorproduct\left({b_{13}} -\frac{{b_{17}}}{2}
 -\frac{{b_{37}}}{2} +\frac{{b_{47}}}{2}\right)\tensorproduct\left({c_{23}} +{c_{32}} -{c_{36}} -{c_{37}} \right),
 \end{aligned}\\
m_{26}=&\begin{aligned}[t]
 \left(-\frac{{a_{12}}}{2}+\frac{{a_{22}}}{2}+{a_{13}} +{a_{24}}
 \right)\tensorproduct\left({b_{45}} -\frac{{b_{17}}}{2}
 +\frac{{b_{37}}}{2} +\frac{{b_{47}}}{2}\right)\tensorproduct\left({c_{11}} +{c_{14}} +{c_{17}} +{c_{25}} \right),
 \end{aligned}\\
m_{27}=&\begin{aligned}[t]
 {a_{32}}\tensorproduct\!\left(\! \frac{{b_{41}}}{2}-\frac{{b_{13}}}{2} -\frac{{b_{11}}}{2}-{b_{21}} -\frac{{b_{31}}}{2} -{b_{23}} +\frac{{b_{33}}}{2}-\frac{{b_{43}}}{2}+\frac{{b_{16}}}{2}-{b_{26}} -\frac{{b_{36}}}{2}-\frac{{b_{46}}}{2}+{b_{27}}\!\right)\!\tensorproduct{c_{37}},
 \end{aligned}\\
m_{28}=&\begin{aligned}[t]
 \left({a_{31}} -\frac{{a_{22}}}{2}+\frac{{a_{32}}}{2}+{a_{23}} \right)\tensorproduct\left({b_{36}} -\frac{{b_{17}}}{2}-\frac{{b_{37}}}{2}+\frac{{b_{47}}}{2}\right)\tensorproduct\left({c_{26}} -{c_{33}} +{c_{35}} -{c_{37}} \right),
 \end{aligned}\\
m_{29}=&\begin{aligned}[t]
 \left({a_{33}} +{a_{34}} \right)\tensorproduct\left({b_{31}} +{b_{43}} +\frac{{b_{17}}}{2}-\frac{{b_{37}}}{2}-\frac{{b_{47}}}{2}\right)\tensorproduct\left({c_{31}} +{c_{33}} +{c_{37}} \right),
 \end{aligned}\\
m_{30}=&\begin{aligned}[t]
 \left({a_{21}} -\frac{{a_{12}}}{2} +\frac{{a_{22}}}{2} +{a_{14}}
 \right)\tensorproduct\left(-{b_{14}} +\frac{{b_{17}}}{2}
 -\frac{{b_{37}}}{2} +\frac{{b_{47}}}{2}\right)\tensorproduct\left(-{c_{12}} +{c_{16}} +{c_{17}} -{c_{24}} \right),
 \end{aligned}\\
m_{31}=&\begin{aligned}[t]
\left(\!\begin{array}{c}\frac{{a_{12}}}{2}+\frac{{a_{32}}}{2}\\+{a_{13}} +{a_{33}}\end{array}\!\right)\!\tensorproduct\!\left(
	\begin{array}{c}
	{b_{23}} +{b_{34}} -{b_{32}} +{b_{25}} -{b_{37}} \\
		+\lfrac{({b_{13}} +{b_{33}}+{b_{43}} -{b_{15}} +{b_{35}}-{b_{45}})}{2}
	\end{array}
	\right)\tensorproduct{c_{13}},
 \end{aligned}\\
m_{32}=&\begin{aligned}[t]
 \left(-{a_{11}} +{a_{21}} +\frac{{a_{12}}}{2}+\frac{{a_{22}}}{2}\right)\tensorproduct\left(\frac{{b_{12}}}{2}+{b_{22}} -\frac{{b_{32}}}{2}-\frac{{b_{42}}}{2}\right)\tensorproduct\left({c_{12}} -{c_{16}} -{c_{17}} +{c_{22}} \right),
 \end{aligned}\\
m_{33}=&\begin{aligned}[t]
 \left(\frac{{a_{22}}}{2}-\frac{{a_{32}}}{2}+{a_{23}} -{a_{33}} \right)\tensorproduct\left(\frac{{b_{13}}}{2}+{b_{23}} +\frac{{b_{33}}}{2}+\frac{{b_{43}}}{2}-{b_{37}} \right)\tensorproduct{c_{23}},
 \end{aligned}\\
m_{34}=&\begin{aligned}[t]
 \left(\frac{{a_{21}}}{2}+\frac{{a_{23}}}{2}\right)\tensorproduct\left({b_{17}} +{b_{37}} -{b_{47}} \right)\tensorproduct\left(-{c_{22}} +{c_{23}} -{c_{25}} +{c_{26}} +{c_{27}} \right),
 \end{aligned}\\
m_{35}=&\begin{aligned}[t]
 \left(-{a_{11}} +\frac{{a_{12}}}{2}+\frac{{a_{32}}}{2}+{a_{33}} \right)\tensorproduct\left({b_{32}} -{b_{13}} +\frac{{b_{17}}}{2}+\frac{{b_{37}}}{2}-\frac{{b_{47}}}{2}\right)\tensorproduct\left({c_{13}} +{c_{32}} \right),
 \end{aligned}\\
m_{36}=&\begin{aligned}[t]
 \left(\frac{{a_{12}}}{2}+\frac{{a_{22}}}{2}-{a_{13}} +{a_{23}} \right)\tensorproduct\left(-\frac{{b_{15}}}{2}+{b_{25}} +\frac{{b_{35}}}{2}-\frac{{b_{45}}}{2}\right)\tensorproduct\left(-{c_{13}} +{c_{15}} -{c_{17}} +{c_{25}} \right),
 \end{aligned}\\
m_{37}=&\begin{aligned}[t]
 \left(\frac{{a_{12}}}{2}-\frac{{a_{22}}}{2}+{a_{13}} -{a_{23}} \right)\tensorproduct\left({b_{32}} -{b_{34}} +\frac{{b_{15}}}{2}-{b_{25}} -\frac{{b_{35}}}{2}+\frac{{b_{45}}}{2}+{b_{37}} \right)\tensorproduct\left({c_{13}} -{c_{15}} +{c_{17}} \right),
 \end{aligned}
\end{align}
\begin{align}
m_{38}=&\begin{aligned}[t]
 \left(\frac{{a_{12}}}{2}+\frac{{a_{22}}}{2}-{a_{14}} +{a_{24}} \right)\tensorproduct\left(-\frac{{b_{14}}}{2}+{b_{24}} -\frac{{b_{34}}}{2}+\frac{{b_{44}}}{2}\right)\tensorproduct\left({c_{11}} +{c_{14}} +{c_{17}} +{c_{24}} \right),
 \end{aligned}\\
m_{39}=&\begin{aligned}[t]
 \left(-\frac{{a_{22}}}{2}+\frac{{a_{32}}}{2}+{a_{33}} +{a_{24}} \right)\tensorproduct\left(-{b_{43}} -\frac{{b_{17}}}{2}+\frac{{b_{37}}}{2}+\frac{{b_{47}}}{2}\right)\tensorproduct\left(-{c_{23}} +{c_{31}} +{c_{34}} +{c_{37}} \right),
 \end{aligned} \\
m_{40}=&\begin{aligned}[t]
 {a_{12}}\tensorproduct\!\left(
	\begin{array}{c}
	{b_{22}} -{b_{24}} +{b_{27}} +{b_{25}} \\
		+\lfrac{({{b_{12}}} -{{b_{32}}}-{{b_{42}}}+{{b_{14}}} +{{b_{34}}}-{{b_{44}}}-{{b_{15}}} +{{b_{35}}}-{{b_{45}}})}{2}
	\end{array}
	\right)\!\tensorproduct{c_{17}},
 \end{aligned}\\
m_{41}=&\begin{aligned}[t]
 \left(-{a_{11}} +\frac{{a_{12}}}{2}\right)\tensorproduct\left(-{b_{12}} +{b_{32}} +{b_{42}} \right)\tensorproduct\left({c_{12}} +{c_{22}} -{c_{32}} \right),
 \end{aligned} \\
m_{42}=&\begin{aligned}[t]
\left(\!\begin{array}{c}\frac{{a_{12}}}{2}+\frac{{a_{32}}}{2}\\+{a_{14}} +{a_{34}}\end{array}\!\right)\!\tensorproduct\!\left(
	\begin{array}{c}
	{b_{21}} -{b_{24}} -{b_{42}} -{b_{45}} -{b_{47}}\\
		+\lfrac{( {{b_{11}}} +{{b_{31}}}+{{b_{41}}} +{{b_{14}}} +{{b_{34}}}-{{b_{44}}})} {2}
	\end{array}
	\right)\tensorproduct{c_{11}},
 \end{aligned}\\
m_{43}=&\begin{aligned}[t]
 \left(\frac{{a_{12}}}{2}-{a_{13}} \right)\tensorproduct\left({b_{15}} -{b_{35}} +{b_{45}} \right)\tensorproduct\left({c_{15}} +{c_{25}} -{c_{35}} \right),
 \end{aligned}\\
m_{44}=&\begin{aligned}[t]
 \left(\frac{{a_{12}}}{2}-{a_{14}} \right)\tensorproduct\left({b_{14}} +{b_{34}} -{b_{44}} \right)\tensorproduct\left({c_{14}} +{c_{24}} -{c_{34}} \right),
 \end{aligned}\\
m_{45}=&\begin{aligned}[t]
\left(\begin{array}{c}-\frac{{a_{12}}}{2} +\frac{{a_{32}}}{2}\\ +{a_{13}} +{a_{33}}\end{array}\right)\tensorproduct\left(\frac{{b_{13}}}{2}+{b_{23}} -\frac{{b_{33}}}{2}+\frac{{b_{43}}}{2}-\frac{{b_{15}}}{2}+{b_{25}} +\frac{{b_{35}}}{2}-\frac{{b_{45}}}{2}\right)\tensorproduct\left(-{c_{13}} +{c_{35}} \right),
 \end{aligned}\\
m_{46}=&\begin{aligned}[t]
 \left({a_{31}} -\frac{{a_{22}}}{2} +\frac{{a_{32}}}{2} +{a_{24}} \right)\tensorproduct\left(-{b_{46}} +\frac{{b_{17}}}{2}-\frac{{b_{37}}}{2}+\frac{{b_{47}}}{2}\right)\tensorproduct\left(-{c_{26}} +{c_{31}} +{c_{34}} +{c_{37}} \right),
 \end{aligned}\\
m_{47}=&\begin{aligned}[t]
 \left(\frac{{a_{22}}}{2} +\frac{{a_{32}}}{2} -{a_{23}} +{a_{33}}
 \right)\tensorproduct\left(\frac{{b_{13}}}{2} +{b_{23}}
 -\frac{{b_{33}}}{2} +\frac{{b_{43}}}{2}\right)\tensorproduct\left({c_{23}} +{c_{33}} -{c_{35}} +{c_{37}} \right),
 \end{aligned}\\
m_{48}=&\begin{aligned}[t]
\left(\!\begin{array}{c}{-a_{11}} -{a_{31}}\\ \frac{{a_{12}}}{2}
 +\frac{{a_{32}}}{2}\end{array}\!\right)\!\tensorproduct\!\left(
	\begin{array}{c}
 {b_{11}} +{b_{13}} -{b_{17}} +{b_{22}} +{b_{26}} \\
			+\lfrac{(
			{{b_{36}}} +{{b_{46}}} -{{b_{12}}} -{{b_{32}}} -{{b_{42}}} -{{b_{16}}}
		)}{2}
	\end{array}
	\right)\tensorproduct{c_{32}},
 \end{aligned}\\
m_{49}=&\begin{aligned}[t]
 \left(-\frac{{a_{12}}}{2} +\frac{{a_{22}}}{2} +{a_{23}} +{a_{14}}
 \right)\tensorproduct\left(-{b_{34}} -\frac{{b_{17}}}{2}
 +\frac{{b_{37}}}{2}
 +\frac{{b_{47}}}{2}\right)\tensorproduct\left({c_{13}} -{c_{15}}
 +{c_{17}} -{c_{24}} \right),
 \end{aligned}\\
m_{50}=&\begin{aligned}[t]
 \left({a_{31}} +\frac{{a_{12}}}{2}+\frac{{a_{32}}}{2}-{a_{14}} \right)\tensorproduct\left({b_{14}} +{b_{46}} -\frac{{b_{17}}}{2}+\frac{{b_{37}}}{2}-\frac{{b_{47}}}{2}\right)\tensorproduct\left(-{c_{16}} +{c_{34}} \right),
 \end{aligned}\\
m_{51}=&\begin{aligned}[t]
 \left(-\frac{{a_{22}}}{2}+\frac{{a_{32}}}{2}+{a_{23}} +{a_{34}} \right)\tensorproduct\left({b_{31}} +\frac{{b_{17}}}{2}-\frac{{b_{37}}}{2}-\frac{{b_{47}}}{2}\right)\tensorproduct\left({c_{21}} -{c_{33}} +{c_{35}} -{c_{37}} \right),
 \end{aligned}\\
m_{52}=&\begin{aligned}[t]
 \left(\frac{{a_{22}}}{2}-\frac{{a_{32}}}{2}+{a_{24}} -{a_{34}} \right)\tensorproduct\left(\frac{{b_{11}}}{2}+{b_{21}} +\frac{{b_{31}}}{2}+\frac{{b_{41}}}{2}-{b_{47}} \right)\tensorproduct{c_{21}},
 \end{aligned}\\
m_{53}=&\begin{aligned}[t]
 \left(\frac{{a_{12}}}{2}-\frac{{a_{22}}}{2}-{a_{13}} +{a_{23}} \right)\tensorproduct\left(\frac{{b_{15}}}{2}-{b_{25}} +\frac{{b_{35}}}{2}+\frac{{b_{45}}}{2}+{b_{37}} \right)\tensorproduct{c_{25}},
 \end{aligned} \\
m_{54}=&\begin{aligned}[t]
\left(\begin{array}{c} \frac{{a_{32}}}{2} -\frac{{a_{12}}}{2}\\ +{a_{14}} +{a_{34}}\end{array}\right)\tensorproduct\left(-\frac{{b_{11}}}{2}-{b_{21}} -\frac{{b_{31}}}{2}+\frac{{b_{41}}}{2}-\frac{{b_{14}}}{2}+{b_{24}} -\frac{{b_{34}}}{2}+\frac{{b_{44}}}{2}\right)\tensorproduct\left({c_{11}} +{c_{34}} \right),
 \end{aligned}\\
m_{55}=&\begin{aligned}[t]
 \left({a_{11}} -\frac{{a_{12}}}{2}+\frac{{a_{22}}}{2}+{a_{24}} \right)\tensorproduct\left({b_{42}} +\frac{{b_{17}}}{2}-\frac{{b_{37}}}{2}+\frac{{b_{47}}}{2}\right)\tensorproduct\left({c_{11}} +{c_{14}} +{c_{17}} +{c_{22}} \right),
 \end{aligned}\\
m_{56}=&\begin{aligned}[t]
 \left(\frac{{a_{12}}}{2}-\frac{{a_{22}}}{2}-{a_{14}} +{a_{24}} \right)\tensorproduct\left(\frac{{b_{14}}}{2}-{b_{24}} +\frac{{b_{34}}}{2}+\frac{{b_{44}}}{2}-{b_{47}} \right)\tensorproduct{c_{24}},
 \end{aligned}\\
m_{57}=&\begin{aligned}[t]
 \left(\frac{{a_{23}}}{2}+\frac{{a_{24}}}{2}\right)\tensorproduct\left(-{b_{17}} +{b_{37}} +{b_{47}} \right)\tensorproduct\left({c_{21}} +{c_{23}} +{c_{24}} -{c_{25}} +{c_{27}} \right),
 \end{aligned}\\
m_{58}=&\begin{aligned}[t]
 \left({a_{21}} -\frac{{a_{22}}}{2}+\frac{{a_{32}}}{2}+{a_{34}} \right)\tensorproduct\left({b_{11}} -\frac{{b_{17}}}{2}+\frac{{b_{37}}}{2}-\frac{{b_{47}}}{2}\right)\tensorproduct\left({c_{21}} +{c_{32}} -{c_{36}} -{c_{37}} \right),
 \end{aligned}\\
m_{59}=&\begin{aligned}[t]
\left(\!\begin{array}{c}{a_{11}} +{a_{31}}\\+\frac{{a_{12}}}{2}+\frac{{a_{32}}}{2}\end{array}\!\right)\!\tensorproduct\!\left(
	{b_{22}} +{b_{14}} -{b_{15}} -{b_{17}} +{b_{26}}
	+\lfrac{(
	{{b_{12}}} -{{b_{32}}}-{{b_{42}}} +{{b_{16}}} +{{b_{36}}}+{{b_{46}}}
	)}{2}
	\right)\tensorproduct{c_{16}},
 \end{aligned}\\
m_{60}=&\begin{aligned}[t]
 \left({a_{31}} +\frac{{a_{12}}}{2}+\frac{{a_{32}}}{2}-{a_{13}} \right)\tensorproduct\left({b_{15}} -{b_{36}} +\frac{{b_{17}}}{2}+\frac{{b_{37}}}{2}-\frac{{b_{47}}}{2}\right)\tensorproduct\left({c_{16}} +{c_{35}} \right),
 \end{aligned}\\
m_{61}=&\begin{aligned}[t]
 \left({a_{31}} +{a_{33}} \right)\tensorproduct\left({b_{13}} +{b_{36}} -\frac{{b_{17}}}{2}-\frac{{b_{37}}}{2}+\frac{{b_{47}}}{2}\right)\tensorproduct\left({c_{33}} +{c_{36}} +{c_{37}} \right),
 \end{aligned}\\
m_{62}=&\begin{aligned}[t]
 \left(\frac{{a_{12}}}{2}-\frac{{a_{22}}}{2}+{a_{14}} -{a_{24}} \right)\tensorproduct\left({b_{42}} -\frac{{b_{14}}}{2}+{b_{24}} -\frac{{b_{34}}}{2}+\frac{{b_{44}}}{2}+{b_{45}} +{b_{47}} \right)\tensorproduct\left({c_{11}} +{c_{14}} +{c_{17}} \right),
 \end{aligned}\\
m_{63}=& {a_{22}}\tensorproduct{b_{27}}\tensorproduct{c_{27}}.
	\label{eq:lastprod63}
\end{align}
The type of this tensor decomposition is:
\begin{equation}
{6X^{2}Y^{3}Z^{2}+15X^{2}Y^{2}Z^{2}+2XY^{4}Z+6X^{2}YZ^{2}+12XY^{3}Z+12XY^{2}Z+10XYZ}.
\end{equation}
Several other representations of this algorithm are given in the following subsection, in~\cref{sec:source}, and are available in~\cite{Sedoglavic:FMMDB,jgd:2024:plinopt}.
\subsection[LRP representation of the rational 3x4x7:63 algorithm]{LRP representation of the rational \FMMA{3}{4}{7}{63} algorithm}\label{sec:LRPrepresentation63}
We present now the \textsc{lrp} representation of the tensor decomposition given by~\Crefrange{eq:firstprod63}{eq:lastprod63}:\\
\scalebox{.7}{%
\(\mat{L}=\begin{smatrix}
0&0&0&0&1&0&0&1&0&0&0&0\\
1&\frac{1}{2}&0&0&-1&-\frac{1}{2}&0&0&0&0&0&0\\
0&0&0&0&0&-\frac{1}{2}&0&1&0&\frac{1}{2}&0&-1\\
1&0&1&0&0&0&0&0&0&0&0&0\\
0&-\frac{1}{2}&0&1&0&0&0&0&0&-\frac{1}{2}&0&1\\
0&0&0&0&0&0&0&0&1&0&0&1\\
0&0&0&0&0&0&0&0&1&\frac{1}{2}&0&0\\
0&0&0&0&1&\frac{1}{2}&0&0&-1&-\frac{1}{2}&0&0\\
0&0&0&0&0&\frac{1}{2}&0&-1&0&\frac{1}{2}&0&1\\
0&\frac{1}{2}&-1&0&0&0&0&0&0&\frac{1}{2}&0&1\\
0&0&0&0&0&0&0&0&0&\frac{1}{2}&1&0\\
0&0&0&0&0&-\frac{1}{2}&1&0&0&\frac{1}{2}&-1&0\\
0&0&0&0&0&0&0&0&0&\frac{1}{2}&0&1\\
0&-\frac{1}{2}&1&0&1&\frac{1}{2}&0&0&0&0&0&0\\
1&0&0&1&0&0&0&0&0&0&0&0\\
0&0&0&0&-1&\frac{1}{2}&0&0&1&\frac{1}{2}&0&0\\
1&-\frac{1}{2}&0&0&0&0&0&0&1&\frac{1}{2}&0&0\\
0&\frac{1}{2}&-1&0&0&0&0&0&0&\frac{1}{2}&-1&0\\
1&-\frac{1}{2}&0&0&0&\frac{1}{2}&1&0&0&0&0&0\\
0&0&0&0&1&-\frac{1}{2}&0&0&-1&\frac{1}{2}&0&0\\
0&0&1&1&0&0&0&0&0&0&0&0\\
0&\frac{1}{2}&0&-1&0&0&0&0&0&\frac{1}{2}&1&0\\
-1&\frac{1}{2}&0&0&1&-\frac{1}{2}&0&0&0&0&0&0\\
-1&\frac{1}{2}&0&0&0&0&0&0&0&\frac{1}{2}&0&1\\
0&0&0&0&1&-\frac{1}{2}&0&0&0&\frac{1}{2}&1&0\\
0&-\frac{1}{2}&1&0&0&\frac{1}{2}&0&1&0&0&0&0\\
0&0&0&0&0&0&0&0&0&1&0&0\\
0&0&0&0&0&-\frac{1}{2}&1&0&1&\frac{1}{2}&0&0\\
0&0&0&0&0&0&0&0&0&0&1&1\\
0&-\frac{1}{2}&0&1&1&\frac{1}{2}&0&0&0&0&0&0\\
0&\frac{1}{2}&1&0&0&0&0&0&0&\frac{1}{2}&1&0\\
-1&\frac{1}{2}&0&0&1&\frac{1}{2}&0&0&0&0&0&0\\
0&0&0&0&0&\frac{1}{2}&1&0&0&-\frac{1}{2}&-1&0\\
0&0&0&0&\frac{1}{2}&0&\frac{1}{2}&0&0&0&0&0\\
-1&\frac{1}{2}&0&0&0&0&0&0&0&\frac{1}{2}&1&0\\
0&\frac{1}{2}&-1&0&0&\frac{1}{2}&1&0&0&0&0&0\\
0&\frac{1}{2}&1&0&0&-\frac{1}{2}&-1&0&0&0&0&0\\
0&\frac{1}{2}&0&-1&0&\frac{1}{2}&0&1&0&0&0&0\\
0&0&0&0&0&-\frac{1}{2}&0&1&0&\frac{1}{2}&1&0\\
0&1&0&0&0&0&0&0&0&0&0&0\\
-1&\frac{1}{2}&0&0&0&0&0&0&0&0&0&0\\
0&\frac{1}{2}&0&1&0&0&0&0&0&\frac{1}{2}&0&1\\
0&\frac{1}{2}&-1&0&0&0&0&0&0&0&0&0\\
0&\frac{1}{2}&0&-1&0&0&0&0&0&0&0&0\\
0&-\frac{1}{2}&1&0&0&0&0&0&0&\frac{1}{2}&1&0\\
0&0&0&0&0&-\frac{1}{2}&0&1&1&\frac{1}{2}&0&0\\
0&0&0&0&0&\frac{1}{2}&-1&0&0&\frac{1}{2}&1&0\\
-1&\frac{1}{2}&0&0&0&0&0&0&-1&\frac{1}{2}&0&0\\
0&-\frac{1}{2}&0&1&0&\frac{1}{2}&1&0&0&0&0&0\\
0&\frac{1}{2}&0&-1&0&0&0&0&1&\frac{1}{2}&0&0\\
0&0&0&0&0&-\frac{1}{2}&1&0&0&\frac{1}{2}&0&1\\
0&0&0&0&0&\frac{1}{2}&0&1&0&-\frac{1}{2}&0&-1\\
0&\frac{1}{2}&-1&0&0&-\frac{1}{2}&1&0&0&0&0&0\\
0&-\frac{1}{2}&0&1&0&0&0&0&0&\frac{1}{2}&0&1\\
1&-\frac{1}{2}&0&0&0&\frac{1}{2}&0&1&0&0&0&0\\
0&\frac{1}{2}&0&-1&0&-\frac{1}{2}&0&1&0&0&0&0\\
0&0&0&0&0&0&\frac{1}{2}&\frac{1}{2}&0&0&0&0\\
0&0&0&0&1&-\frac{1}{2}&0&0&0&\frac{1}{2}&0&1\\
1&\frac{1}{2}&0&0&0&0&0&0&1&\frac{1}{2}&0&0\\
0&\frac{1}{2}&-1&0&0&0&0&0&1&\frac{1}{2}&0&0\\
0&0&0&0&0&0&0&0&1&0&1&0\\
0&\frac{1}{2}&0&1&0&-\frac{1}{2}&0&-1&0&0&0&0\\
0&0&0&0&0&1&0&0&0&0&0&0\\
\end{smatrix},
\mat{R}=\begin{smatrix}
0&0&0&0&0&0&1&0&0&0&0&0&0&0&0&0&0&0&0&0&-1&0&0&0&0&0&0&1\\
0&-\frac{1}{2}&0&-1&1&0&1&0&-1&0&0&0&0&0&0&\frac{1}{2}&0&0&0&0&0&0&\frac{1}{2}&0&0&0&0&0\\
\frac{1}{2}&0&0&0&0&0&0&1&0&0&0&0&0&0&\frac{1}{2}&0&0&0&0&0&0&-\frac{1}{2}&0&1&0&0&1&-1\\
0&0&0&0&1&0&\frac{1}{2}&0&0&0&0&0&0&0&0&1&0&0&0&0&\frac{1}{2}&0&0&0&0&0&0&-\frac{1}{2}\\
\frac{1}{2}&0&0&\frac{1}{2}&0&0&0&1&0&0&-1&0&0&0&\frac{1}{2}&0&0&\frac{1}{2}&0&0&0&-\frac{1}{2}&0&1&\frac{1}{2}&0&1&-1\\
1&0&0&0&0&0&-\frac{1}{2}&0&0&0&0&0&0&0&0&0&0&0&0&0&\frac{1}{2}&0&0&0&0&0&1&-\frac{1}{2}\\
0&0&0&0&0&1&0&0&0&0&0&0&0&0&0&0&0&0&0&-1&0&0&0&0&0&0&-1&0\\
0&0&0&0&0&\frac{1}{2}&-1&0&0&0&0&0&1&0&0&0&0&0&0&\frac{1}{2}&0&0&0&0&0&0&\frac{1}{2}&0\\
\frac{1}{2}&0&0&0&0&0&0&1&0&0&0&0&0&0&\frac{1}{2}&0&0&0&0&0&0&-\frac{1}{2}&0&0&0&0&0&0\\
0&0&0&0&0&0&-\frac{1}{2}&0&0&0&0&0&0&0&-1&0&0&0&0&0&\frac{1}{2}&0&0&0&0&1&0&\frac{1}{2}\\
0&0&1&0&0&0&0&0&0&0&0&0&0&0&0&0&-1&0&0&0&0&0&0&1&0&0&0&0\\
0&0&\frac{1}{2}&0&0&0&0&0&0&1&0&0&0&0&1&0&-\frac{1}{2}&0&0&1&-1&0&0&\frac{1}{2}&0&0&0&0\\
1&0&0&0&0&0&0&0&0&0&0&0&0&0&1&0&0&0&0&0&0&-1&0&0&0&0&0&0\\
0&0&0&0&1&0&\frac{1}{2}&0&0&0&0&0&0&0&0&0&0&0&0&0&\frac{1}{2}&0&0&0&0&0&0&-\frac{1}{2}\\
0&0&0&-1&0&0&\frac{1}{2}&0&0&0&0&0&0&0&0&0&0&0&0&0&-\frac{1}{2}&0&1&0&0&0&0&\frac{1}{2}\\
0&0&0&0&0&-\frac{1}{2}&0&0&0&0&0&0&1&0&0&0&0&0&0&\frac{1}{2}&0&0&0&0&0&0&\frac{1}{2}&0\\
0&\frac{1}{2}&0&0&0&-\frac{1}{2}&0&0&1&0&0&0&1&0&0&-\frac{1}{2}&0&0&0&\frac{1}{2}&0&0&-\frac{1}{2}&0&0&0&\frac{1}{2}&0\\
0&0&\frac{1}{2}&0&-\frac{1}{2}&0&0&0&0&1&0&1&0&0&1&0&-\frac{1}{2}&0&-\frac{1}{2}&1&-1&0&0&\frac{1}{2}&0&-\frac{1}{2}&0&0\\
0&0&0&0&0&0&\frac{1}{2}&0&0&0&0&0&0&0&0&1&0&0&0&0&\frac{1}{2}&0&0&0&0&0&0&-\frac{1}{2}\\
1&0&1&0&0&-\frac{1}{2}&-1&0&0&0&0&0&1&0&0&0&0&0&0&\frac{1}{2}&0&0&0&0&0&0&\frac{1}{2}&0\\
0&0&0&0&0&0&-\frac{1}{2}&0&0&0&0&0&0&0&0&0&0&-1&0&0&\frac{1}{2}&0&0&0&0&1&0&\frac{1}{2}\\
0&0&0&0&0&0&\frac{1}{2}&0&0&0&0&0&0&0&0&0&0&1&0&0&-\frac{1}{2}&0&0&1&0&0&0&-\frac{1}{2}\\
0&\frac{1}{2}&0&0&0&0&1&0&-1&0&0&0&0&0&0&\frac{1}{2}&0&0&0&0&0&0&\frac{1}{2}&0&0&0&0&0\\
-1&0&0&0&0&0&\frac{1}{2}&0&0&0&0&0&0&0&0&0&0&0&0&0&-\frac{1}{2}&0&1&0&0&0&0&\frac{1}{2}\\
0&0&1&0&0&0&-\frac{1}{2}&0&0&0&0&0&0&0&0&0&0&0&0&0&-\frac{1}{2}&0&0&0&0&0&0&\frac{1}{2}\\
0&0&0&0&0&0&-\frac{1}{2}&0&0&0&0&0&0&0&0&0&0&0&0&0&\frac{1}{2}&0&0&0&0&1&0&\frac{1}{2}\\
-\frac{1}{2}&0&-\frac{1}{2}&0&0&\frac{1}{2}&0&-1&0&-1&0&0&-1&1&-\frac{1}{2}&0&\frac{1}{2}&0&0&-\frac{1}{2}&0&\frac{1}{2}&0&-\frac{1}{2}&0&0&-\frac{1}{2}&0\\
0&0&0&0&0&0&-\frac{1}{2}&0&0&0&0&0&0&0&0&0&0&0&0&1&-\frac{1}{2}&0&0&0&0&0&0&\frac{1}{2}\\
0&0&0&0&0&0&\frac{1}{2}&0&0&0&0&0&0&0&1&0&0&0&0&0&-\frac{1}{2}&0&0&1&0&0&0&-\frac{1}{2}\\
0&0&0&-1&0&0&\frac{1}{2}&0&0&0&0&0&0&0&0&0&0&0&0&0&-\frac{1}{2}&0&0&0&0&0&0&\frac{1}{2}\\
0&0&\frac{1}{2}&0&-\frac{1}{2}&0&0&0&0&1&0&1&0&0&0&-1&\frac{1}{2}&1&\frac{1}{2}&0&-1&0&0&\frac{1}{2}&0&-\frac{1}{2}&0&0\\
0&\frac{1}{2}&0&0&0&0&0&0&1&0&0&0&0&0&0&-\frac{1}{2}&0&0&0&0&0&0&-\frac{1}{2}&0&0&0&0&0\\
0&0&\frac{1}{2}&0&0&0&0&0&0&1&0&0&0&0&0&0&\frac{1}{2}&0&0&0&-1&0&0&\frac{1}{2}&0&0&0&0\\
0&0&0&0&0&0&1&0&0&0&0&0&0&0&0&0&0&0&0&0&1&0&0&0&0&0&0&-1\\
0&0&-1&0&0&0&\frac{1}{2}&0&0&0&0&0&0&0&0&1&0&0&0&0&\frac{1}{2}&0&0&0&0&0&0&-\frac{1}{2}\\
0&0&0&0&-\frac{1}{2}&0&0&0&0&0&0&1&0&0&0&0&0&0&\frac{1}{2}&0&0&0&0&0&0&-\frac{1}{2}&0&0\\
0&0&0&0&\frac{1}{2}&0&0&0&0&0&0&-1&0&0&0&1&0&-1&-\frac{1}{2}&0&1&0&0&0&0&\frac{1}{2}&0&0\\
0&0&0&-\frac{1}{2}&0&0&0&0&0&0&1&0&0&0&0&0&0&-\frac{1}{2}&0&0&0&0&0&0&\frac{1}{2}&0&0&0\\
0&0&0&0&0&0&-\frac{1}{2}&0&0&0&0&0&0&0&0&0&0&0&0&0&\frac{1}{2}&0&0&-1&0&0&0&\frac{1}{2}\\
0&\frac{1}{2}&0&\frac{1}{2}&-\frac{1}{2}&0&0&0&1&0&-1&1&0&1&0&-\frac{1}{2}&0&\frac{1}{2}&\frac{1}{2}&0&0&0&-\frac{1}{2}&0&-\frac{1}{2}&-\frac{1}{2}&0&0\\
0&-1&0&0&0&0&0&0&0&0&0&0&0&0&0&1&0&0&0&0&0&0&1&0&0&0&0&0\\
\frac{1}{2}&0&0&\frac{1}{2}&0&0&0&1&0&0&-1&0&0&0&\frac{1}{2}&0&0&\frac{1}{2}&0&0&0&\frac{1}{2}&-1&0&-\frac{1}{2}&-1&0&-1\\
0&0&0&0&1&0&0&0&0&0&0&0&0&0&0&0&0&0&-1&0&0&0&0&0&0&1&0&0\\
0&0&0&1&0&0&0&0&0&0&0&0&0&0&0&0&0&1&0&0&0&0&0&0&-1&0&0&0\\
0&0&\frac{1}{2}&0&-\frac{1}{2}&0&0&0&0&1&0&1&0&0&0&0&-\frac{1}{2}&0&\frac{1}{2}&0&0&0&0&\frac{1}{2}&0&-\frac{1}{2}&0&0\\
0&0&0&0&0&0&\frac{1}{2}&0&0&0&0&0&0&0&0&0&0&0&0&0&-\frac{1}{2}&0&0&0&0&0&-1&\frac{1}{2}\\
0&0&\frac{1}{2}&0&0&0&0&0&0&1&0&0&0&0&0&0&-\frac{1}{2}&0&0&0&0&0&0&\frac{1}{2}&0&0&0&0\\
1&-\frac{1}{2}&1&0&0&-\frac{1}{2}&-1&0&1&0&0&0&1&0&0&-\frac{1}{2}&0&0&0&\frac{1}{2}&0&0&-\frac{1}{2}&0&0&0&\frac{1}{2}&0\\
0&0&0&0&0&0&-\frac{1}{2}&0&0&0&0&0&0&0&0&0&0&-1&0&0&\frac{1}{2}&0&0&0&0&0&0&\frac{1}{2}\\
0&0&0&1&0&0&-\frac{1}{2}&0&0&0&0&0&0&0&0&0&0&0&0&0&\frac{1}{2}&0&0&0&0&0&1&-\frac{1}{2}\\
0&0&0&0&0&0&\frac{1}{2}&0&0&0&0&0&0&0&1&0&0&0&0&0&-\frac{1}{2}&0&0&0&0&0&0&-\frac{1}{2}\\
\frac{1}{2}&0&0&0&0&0&0&1&0&0&0&0&0&0&\frac{1}{2}&0&0&0&0&0&0&\frac{1}{2}&0&0&0&0&0&-1\\
0&0&0&0&\frac{1}{2}&0&0&0&0&0&0&-1&0&0&0&0&0&0&\frac{1}{2}&0&1&0&0&0&0&\frac{1}{2}&0&0\\
-\frac{1}{2}&0&0&-\frac{1}{2}&0&0&0&-1&0&0&1&0&0&0&-\frac{1}{2}&0&0&-\frac{1}{2}&0&0&0&\frac{1}{2}&0&0&\frac{1}{2}&0&0&0\\
0&0&0&0&0&0&\frac{1}{2}&0&0&0&0&0&0&0&0&0&0&0&0&0&-\frac{1}{2}&0&1&0&0&0&0&\frac{1}{2}\\
0&0&0&\frac{1}{2}&0&0&0&0&0&0&-1&0&0&0&0&0&0&\frac{1}{2}&0&0&0&0&0&0&\frac{1}{2}&0&0&-1\\
0&0&0&0&0&0&-1&0&0&0&0&0&0&0&0&0&0&0&0&0&1&0&0&0&0&0&0&1\\
1&0&0&0&0&0&-\frac{1}{2}&0&0&0&0&0&0&0&0&0&0&0&0&0&\frac{1}{2}&0&0&0&0&0&0&-\frac{1}{2}\\
0&\frac{1}{2}&0&1&-1&\frac{1}{2}&-1&0&1&0&0&0&1&0&0&-\frac{1}{2}&0&0&0&\frac{1}{2}&0&0&-\frac{1}{2}&0&0&0&\frac{1}{2}&0\\
0&0&0&0&1&0&\frac{1}{2}&0&0&0&0&0&0&0&0&0&0&0&0&-1&\frac{1}{2}&0&0&0&0&0&0&-\frac{1}{2}\\
0&0&1&0&0&0&-\frac{1}{2}&0&0&0&0&0&0&0&0&0&0&0&0&1&-\frac{1}{2}&0&0&0&0&0&0&\frac{1}{2}\\
0&0&0&-\frac{1}{2}&0&0&0&0&0&0&1&0&0&0&0&0&0&-\frac{1}{2}&0&0&0&0&1&0&\frac{1}{2}&1&0&1\\
0&0&0&0&0&0&0&0&0&0&0&0&0&1&0&0&0&0&0&0&0&0&0&0&0&0&0&0\\
\end{smatrix}
\)
}\\[10pt]
\scalebox{0.66}{%
\(\mat{P}=\begin{smatrix}
0&0&0&0&0&0&0&0&0&1&0&0&1&0&0&0&0&0&0&0&0&0&0&1&0&1&0&0&0&0&0&0&0&0&0&0&0&1&0&0&0&1&0&0&0&0&0&0&0&0&0&0&0&1&1&0&0&0&0&0&0&1&0\\
0&-1&0&1&0&0&0&0&0&0&0&0&0&-1&1&0&0&0&0&0&0&0&0&0&0&0&0&0&0&-1&0&1&0&0&0&0&0&0&0&0&1&0&0&0&0&0&0&0&0&0&0&0&0&0&0&0&0&0&0&0&0&0&0\\
0&0&0&0&0&0&0&0&0&0&1&0&0&0&0&0&0&0&1&0&0&-1&0&0&0&0&0&0&0&0&1&0&0&0&1&-1&1&0&0&0&0&0&0&0&-1&0&0&0&1&0&0&0&0&0&0&0&0&0&0&0&0&0&0\\
0&0&0&0&0&0&0&0&0&0&0&0&0&0&-1&0&0&0&0&0&-1&0&0&0&0&1&0&0&0&0&0&0&0&0&0&0&0&1&0&0&0&0&0&1&0&0&0&0&0&0&0&0&0&0&1&0&0&0&0&0&0&1&0\\
0&0&0&1&0&0&0&0&0&0&0&0&0&0&0&0&0&0&-1&0&1&0&0&0&0&0&0&0&0&0&0&0&0&0&0&1&-1&0&0&0&0&0&1&0&0&0&0&0&-1&0&0&0&0&0&0&0&0&0&0&0&0&0&0\\
0&1&0&0&0&0&-1&0&0&0&0&0&0&1&0&0&-1&0&0&0&0&0&0&0&0&0&0&0&0&1&0&-1&0&0&0&0&0&0&0&0&0&0&0&0&0&0&0&0&0&-1&0&0&0&0&0&0&0&0&1&1&0&0&0\\
0&1&0&-1&0&0&0&0&0&0&0&0&0&1&-1&0&0&0&1&0&-1&0&0&0&0&1&0&0&0&1&0&-1&0&0&0&-1&1&1&0&1&0&0&0&0&0&0&0&0&1&0&0&0&0&0&1&0&0&0&0&0&0&1&0\\
\frac{1}{2}&0&0&0&0&0&0&0&1&0&0&0&-1&0&0&0&0&0&0&0&0&0&0&0&0&0&0&0&0&0&0&0&0&0&0&0&0&0&0&0&0&0&0&0&0&0&0&0&0&0&1&1&0&0&0&0&1&1&0&0&0&0&0\\
-\frac{1}{2}&0&0&0&0&0&0&0&0&0&0&0&0&0&0&0&0&0&1&0&0&0&1&0&0&0&0&0&0&0&0&1&0&-1&0&0&0&0&0&0&1&0&0&0&0&0&0&0&0&0&0&0&0&0&1&0&0&0&0&0&0&0&0\\
0&0&0&0&0&0&0&0&0&0&-1&0&0&0&0&0&0&0&0&0&0&0&0&0&1&0&0&0&0&0&0&0&1&1&0&0&0&0&-1&0&0&0&0&0&0&0&1&0&0&0&0&0&0&0&0&0&1&0&0&0&0&0&0\\
\frac{1}{2}&0&0&0&0&0&0&0&0&0&0&0&0&0&0&0&0&0&0&0&0&0&0&0&0&0&0&0&0&-1&0&0&0&0&0&0&0&1&0&0&0&0&0&1&0&0&0&0&-1&0&0&0&0&0&0&1&1&0&0&0&0&0&0\\
0&0&0&0&0&0&0&0&0&0&0&0&0&1&0&0&0&0&0&0&0&0&0&0&0&1&0&0&0&0&0&0&0&-1&0&1&0&0&0&0&0&0&1&0&0&0&0&0&0&0&0&0&1&0&0&0&-1&0&0&0&0&0&0\\
\frac{1}{2}&0&0&0&0&0&1&1&0&0&0&0&0&0&0&1&0&0&0&0&0&0&0&0&0&0&0&1&0&0&0&0&0&1&0&0&0&0&0&0&0&0&0&0&0&-1&0&0&0&0&0&0&0&0&0&0&0&0&0&0&0&0&0\\
\frac{1}{2}&0&0&0&0&0&0&0&0&0&0&0&0&0&0&0&0&0&0&0&0&0&0&0&0&0&0&0&0&0&0&0&0&1&0&0&0&0&0&0&0&0&0&0&0&0&0&0&0&0&0&0&0&0&0&0&1&0&0&0&0&0&1\\
0&0&1&0&0&1&0&0&1&0&0&0&-1&0&0&0&0&0&0&0&0&0&0&0&0&0&0&0&1&0&0&0&0&0&0&0&0&0&1&0&0&0&0&0&0&1&0&0&0&0&0&0&0&0&0&0&0&0&0&0&0&0&0\\
0&0&0&0&0&0&0&0&0&0&0&0&0&0&0&-1&1&0&0&-1&0&0&0&1&1&0&0&0&0&0&0&0&0&0&1&0&0&0&0&0&-1&0&0&0&0&0&0&1&0&0&0&0&0&0&0&0&0&1&0&0&0&0&0\\
0&0&0&0&0&0&0&0&0&0&-1&1&0&0&0&0&0&0&0&0&0&0&0&0&0&0&0&-1&1&0&0&0&0&0&0&0&0&0&0&0&0&0&0&0&0&0&1&0&0&0&-1&0&0&0&0&0&0&0&0&0&1&0&0\\
0&0&1&0&1&0&0&0&1&0&0&0&0&0&0&0&0&0&0&0&0&1&0&0&0&0&0&0&0&0&0&0&0&0&0&0&0&0&1&0&0&0&0&-1&0&1&0&0&0&1&0&0&0&1&0&0&0&0&0&0&0&0&0\\
0&0&0&0&0&0&0&0&0&1&0&-1&0&0&0&0&0&1&0&0&0&0&0&0&0&0&0&1&0&0&0&0&0&0&0&0&0&0&0&0&0&0&-1&0&1&0&-1&0&0&0&1&0&0&0&0&0&0&0&0&1&0&0&0\\
0&0&0&0&0&1&1&0&0&0&0&0&0&0&0&1&0&0&0&1&0&0&0&0&-1&0&0&0&0&0&0&0&0&0&0&0&0&0&0&0&0&0&0&0&0&0&0&0&0&0&0&0&0&0&0&0&0&-1&0&0&1&0&0\\
0&0&1&0&0&1&0&0&1&0&0&1&0&0&0&1&0&0&0&1&0&0&0&0&-1&0&1&-1&1&0&0&0&0&0&0&0&0&0&1&0&0&0&0&0&0&1&1&0&0&0&-1&0&0&0&0&0&0&-1&0&0&1&0&0\\
\end{smatrix}
\)
}
\subsection[Straight-line programs for the rational 3x4x7:63 algorithm]{Straight-line programs for the rational~\(\FMMA{3}{4}{7}{63}\)
 algorithm}\label{sec:NumericalSchemeAndComplexity63}
We now give some (\plinopt~generated~\cite{jgd:2024:plinopt}) straight-line programs obtained from~\({\mat{L},\mat{R},\mat{P}}\) matrices presented in~\Cref{sec:LRPrepresentation63} and associated to the tensor decomposition given by~\Crefrange{eq:firstprod63}{eq:lastprod63}:
\par
\begin{lstlisting}[style=slp,caption={\SLP for input linear forms defined by matrix~\(\mat{L}\)},label=lst:347L]
x26:=A32-A31; w76:=A21+A22; w83:=A21-A22; l40:=A12-A11; x58:=A12+A11; l0:=A21+A24; l22:=w83+l40; l12:=A32+A34; l33:=A21+A23; l56:=A24+A23; l42:=A12-A13; w32:=A12+A13; l43:=A12-A14; w36:=A12+A14; e1:=l33-w83; w59:=l33-w76; l31:=l40+w76; e2:=l0-w76; e3:=l0-w83; l6:=A32+A31; l13:=w76-l42; l52:=l42+w59; l60:=A31+A33; l10:=A32+A33; l28:=A34+A33; l9:=l42+l12; l57:=w83+l12; l50:=w59+l12; l23:=l40+l12; l29:=w76-l43; l53:=l12-l43; l55:=l43+e2; l8:=l12-e2; l45:=e2+l6; l49:=l43+l6; l7:=w76-l6; l15:=l6-w83; l16:=l6-l40; l59:=l42+l6; l27:=w59+l6; l38:=e2+l10; l21:=l43+l10; l24:=w83+l10; l34:=l40+l10; l44:=l10-l42; l46:=l10-w59; l2:=l38-l28; l5:=l45-l2; l4:=l28-l21; l17:=l59-l60; l11:=l27-l60; l47:=l40+x26; l19:=w83+x26; l51:=e3-l12; l25:=e3-l42; l37:=l43+e3; l14:=w36-l40; l41:=l12+w36; l61:=w36-e3; l30:=l10+w32; l3:=w32-l40; l20:=w32-l43; l36:=w32-e1; l35:=l42+e1; l48:=e1-l43; l32:=e1-l10; l18:=l33-l22; l54:=l0-l22; l1:=x58-w76; l58:=l6+x58; l26:=A32; l39:=A12; l62:=A22;
\end{lstlisting}
\begin{lstlisting}[style=slp,caption={\SLP for input linear forms defined by matrix~\(\mat{R}\)},label=lst:347R]
r33:=(B17-B47+B37)/2; r0:=B17-r33; r56:=B47-r0; r57:=B11-r0; r12:=B11+B31-B41; r18:=r33+B32; r54:=r0+B42; r40:=B32+B42-B12; r13:=r33+B15; r42:=B15+B45-B35; r24:=B13-r33; r10:=B13+B43-B33; e9:=B23+r10; r46:=e9/2; j5:=r46-B37; r32:=B33+j5; e11:=B25-r42; r35:=e11/2; g9:=B35-r35; r44:=r46+r35; e7:=B21+r12; r8:=e7/2; d51:=B41+r8; e8:=B22-r40; r31:=e8/2; j8:=r31-B12; j9:=B15-r31-B14; r29:=r0-B14; j10:=r35-B32+B34; r43:=B14+B34-B44; e10:=B24-r43; r37:=e10/2; j12:=r37+B47; r53:=r37-r8; r55:=B44-j12; r61:=B42+B45+j12; r27:=B36-r33; r11:=B31+j5+B36; r6:=B16-B36-B46; r45:=r0-B46; j14:=B43+r8+B46; e12:=B26-r6; r15:=e12/2; j16:=r15-B17; r16:=r31+r15; r19:=B11+B13+j16; r7:=B16+j16; r48:=r56-B34; r50:=B31-r56; r25:=B45+r56; r38:=r56-B43; r1:=B17+j9; r2:=j14-B47; r3:=B15+r18; r34:=r18-B13; r4:=j14+r55; r5:=B11-r45; r49:=B14-r45; r9:=B45-r50; r28:=B43+r50; r14:=r54-B14; r23:=r54-B11; r52:=B37+g9; r17:=r11-g9; r20:=B45+r48; r21:=B43-r48; r22:=B17-j8; r26:=B27-e9-e7-e12; r39:=e11+e8-e10+B27; r36:=B37-j10; r30:=r32+j10; r41:=d51-r61; r51:=d51-B47; r47:=j8+r19; r58:=r7-j9; r59:=B15-r27; r60:=B13+r27; r62:=B27;
\end{lstlisting}
\begin{lstlisting}[style=slp, caption={Products},label=lst:347H]
p0:=l0*r0; p1:=l1*r1; p2:=l2*r2; p3:=l3*r3; p4:=l4*r4; p5:=l5*r5; p6:=l6*r6; p7:=l7*r7; p8:=l8*r8; p9:=l9*r9; p10:=l10*r10; p11:=l11*r11; p12:=l12*r12; p13:=l13*r13; p14:=l14*r14; p15:=l15*r15; p16:=l16*r16; p17:=l17*r17; p18:=l18*r18; p19:=l19*r19; p20:=l20*r20; p21:=l21*r21; p22:=l22*r22; p23:=l23*r23; p24:=l24*r24; p25:=l25*r25; p26:=l26*r26; p27:=l27*r27; p28:=l28*r28; p29:=l29*r29; p30:=l30*r30; p31:=l31*r31; p32:=l32*r32; p33:=l33*r33; p34:=l34*r34; p35:=l35*r35; p36:=l36*r36; p37:=l37*r37; p38:=l38*r38; p39:=l39*r39; p40:=l40*r40; p41:=l41*r41; p42:=l42*r42; p43:=l43*r43; p44:=l44*r44; p45:=l45*r45; p46:=l46*r46; p47:=l47*r47; p48:=l48*r48; p49:=l49*r49; p50:=l50*r50; p51:=l51*r51; p52:=l52*r52; p53:=l53*r53; p54:=l54*r54; p55:=l55*r55; p56:=l56*r56; p57:=l57*r57; p58:=l58*r58; p59:=l59*r59; p60:=l60*r60; p61:=l61*r61; p62:=l62*r62;
\end{lstlisting}
\begin{lstlisting}[style=slp,caption={\SLP for input linear forms defined by matrix~\(\mat{P}\)},label=lst:347P]
q12:=p56+p33; q20:=p61+p54+p37+p25; q29:=p48+p36-p35+p18; q34:=p57+p24-p19-p15; q40:=p50-p46+p27-p11; q53:=p45+p38+p8+p2; q54:=p29-p31+p13+p1; z36:=p56+p0; q55:=p33+p0; C24:=p55-p48+p43+p37-p29+z36; C21:=p57+p51+p50-p12+p8+z36; C25:=p52+p42+p35+p25+p13-q12; C26:=p27-p45+p15+p7+p6+q55; C22:=p54+p40+p31+p22+p18-q55; C27:=p62+p0+q12; C23:=p46-p38+p32+p24-p10+q12; z23:=p5+q53; C34:=p53+p49-p43+p21+p4+q53; q56:=q20-p20-p14; C11:=p53+p41+p23+p12+p9+q20; q57:=p60+p28-q40; C35:=p59+p44-p42+p17+p9+q40; C32:=p47-p40+p34+p23+p16+q34; q58:=q54-p3; C16:=p59+p58-p49-p16-p6+q54; C36:=p60+p6+p5-q34; C15:=p42+p20+p3-q29; C13:=p34-p44+p30-p21+p10+q29; C12:=p40+p14-q58; C31:=p28-p12+z23; C14:=p43+q56; C17:=p39+q29+q58+q56; C33:=q57-p10; C37:=p26-q34+z23+q57;
\end{lstlisting}
\par
The program for matrix multiplications
combining~\cref{lst:347L,lst:347R,lst:347H,lst:347P}  together with a Maple command for its
verification is embedded within this~\textsc{pdf} article as the text file
\verb!3x4x7_63_check.mpl!\embedfile{3x4x7_63_check.mpl}.
These straight-line programs require:
\begin{itemize}
\item \(70\) additions for \texttt{L};
\item \(88\) additions and~\(7\) divisions (binary shifts) for \texttt{R};
\item \(107\) additions for \texttt{P}.
\end{itemize}
One could similarly compute the leading constant of asymptotic complexity of the~\(\FMMA{3}{4}{7}{63}\) tensor decomposition, using the operation counts given above, but this requires considering square matrices and the natural candidate for this study would be the tensor decomposition~\(\FMMA{84}{84}{84}{250057}\).
In fact, since~\(84\) is equal to~\({{3}\cdot{4}\cdot{7}}\) and~\(250057\) is equal to~\(63^{3}\), such a tensor decomposition is obtained by the tensor product of~\(\FMMA{3}{4}{7}{63}\) with~\(\FMMA{7}{3}{4}{63}\) and~\(\FMMA{4}{7}{3}{63}\).
Up to Tellegen's transposition principle, the other two tensor decompositions would share the same~\({\mat{L},\mat{R},\mat{P}}\) matrices (permuted and transposed) and thus similar \SLPs.
This would then give an algorithm with cost~\({\GO{n^{3\log_{84}{\!63}}}}\), that is~\({\GO{n^{2.80522}}}\).
\par
Nevertheless, as shown in~\cite{Schwartz:2025aa}, there exists a tensor decomposition~\(\FMMA{84}{84}{84}{224890}\) with a better asymptotic cost:~\({\GO{n^{\log_{84}{\!224890}}}}\) that is~\({\GO{n^{2.7812856}}}\).
\subsection[Alternative bases for the 3x4x7:63 algorithm]{Alternative bases for the \FMMA{3}{4}{7}{63} algorithm}\label{sec:AlternativeBasis63}
Following~\cite{Karstadt:2017aa,Beniamini:2019aa}, we present in this section an alternative basis derived from the tensor decomposition given by~\Crefrange{eq:firstprod63}{eq:lastprod63}.
We choose factorizations of the~\({\mat{L},\mat{R},\mat{P}}\) matrices into matrix products~\({\mat{X}=\mat{X}_{\textup{alt}} \cdot \mat{X}_{\textup{cob}}}\) with common inner dimension~\(62\).
Still with~\(0\)-based indexing,
and looking at the \SLP of \cref{lst:347L,lst:347R},
we see that \texttt{l54:=l0-l22} and that
\texttt{r53:=r37-r8}.
We see also that there is a linear combination of~\(2\) columns
of~\(\mat{P}\) that gives a third one:~\({c_{25}=c_{52}+c_{61}}\).
Thus up to permutations,
we can define
\({\mat{L}_{\textup{alt}}=\begin{smatrix}\mat{I}_{62}\\\vec{e_0}+\vec{e_{22}}\end{smatrix}}\)
and
\({\mat{L}_{\textup{cob}}}\) as~\(\mat{L}\) with its~\(54\)-th row
removed;
\({\mat{R}_{\textup{alt}}=\begin{smatrix}\mat{I}_{62}\\\vec{e_{37}}-\vec{e_{8}}\end{smatrix}}\)
and
\({\mat{R}_{\textup{cob}}}\) as~\(\mat{R}\) with its~\(53\)-th row
removed;
\({\mat{P}_{\textup{alt}}=\Transpose{\begin{smatrix}\mat{I}_{62}\\\vec{e_{52}}+\vec{e_{61}}\end{smatrix}}}\)
and~\(\mat{P}_{\textup{cob}}\) as~\(\mat{P}\)
with its~\(25\)-th column removed.

The algorithm defined by
these~\({\mat{L}_{\textup{alt}},\mat{R}_{\textup{alt}},\mat{P}_{\textup{alt}}}\)
matrices, can be realized with straight-line programs with,
respectively,~\({1,1}\) and~\(2\) additions (for the latter, by the
transposition principle).
Finally, the straight-line programs associated to
$\mat{L}_{\textup{cob}}$,
$\mat{R}_{\textup{cob}}$ and $\mat{P}_{\textup{cob}}$, require:
\begin{itemize}
\item \(69\) additions for~\(\mat{L}_{\textup{cob}}\);
\item \(87\) additions and~\(7\) multiplications (binary shifts) for~\(\mat{R}_{\textup{cob}}\);
\item \(105\) additions for~\(\mat{P}_{\textup{cob}}\).
\end{itemize}
\end{document}